\documentclass[osajnl,twocolumn,showpacs,superscriptaddress,10pt]{revtex4-1}
\usepackage{amsmath,amssymb,graphicx}
\usepackage{bm}

\usepackage{esint}
\usepackage{siunitx}

\usepackage{graphicx}
\usepackage{dcolumn}
\usepackage{bm}
\usepackage{color}
\usepackage{times}
\usepackage{wasysym}
\usepackage{times}
\usepackage{multirow}
\usepackage{verbatim}

\newcommand{\vecr}{\ensuremath{\mathbf{r}}}

\newcommand{\w}{\ensuremath{\omega}}
\newcommand{\e}{\ensuremath{\epsilon}}
\newcommand{\epsbar}{\ensuremath{\bar{\epsilon}}}
\newcommand{\lp}{\ensuremath{\left(}}
\newcommand{\rp}{\ensuremath{\right)}}

\DeclareMathAlphabet{\mathbfit}{OML}{cmm}{b}{it}

\renewcommand{\vec}[1]{\mathbf{#1}}

\newcommand{\figref}[1]{Fig.~\ref{fig:#1}}

\newcommand{\Tabref}[1]{Table~\ref{tab:#1}}
\newcommand{\Figref}[1]{Figure~\ref{fig:#1}}

\renewcommand{\eqref}[1]{(\ref{eq:#1})}

\newcommand{\eqreftwo}[2]{Eqs.~(\ref{eq:#1}) and (\ref{eq:#2})}

\renewcommand{\vec}[1]{\mathbf{#1}}

\renewcommand{\vec}[1]{\mathbf{#1}}

\newcommand{\vE}{\ensuremath{\mathbf{E}}}

\begin{document}


\title{Inverse-designed photonic fibers and metasurfaces for nonlinear
  frequency conversion}

\author{Chawin Sitawarin}
\affiliation{Department of Electrical Engineering, Princeton University, Princeton, NJ, 08544}
\author{Weiliang Jin}
\affiliation{Department of Electrical Engineering, Princeton University, Princeton, NJ, 08544}
\author{Zin Lin}
\affiliation{John A. Paulson School of Engineering and Applied Sciences, Harvard University, Cambridge, MA 02138}
\author{Alejandro W. Rodriguez}
\affiliation{Department of Electrical Engineering, Princeton University, Princeton, NJ, 08544}
\email{arod@princeton.edu}

\begin{abstract}
  Typically, photonic waveguides designed for nonlinear frequency
  conversion rely on intuitive and established principles, including
  index guiding and band-gap engineering, and are based on simple
  shapes with high degrees of symmetry. We show that recently
  developed inverse-design techniques can be applied to discover new
  kinds of microstructured fibers and metasurfaces designed to achieve
  large nonlinear frequency-conversion efficiencies. As a proof of
  principle, we demonstrate complex, wavelength-scale
  chalcogenide--glass fibers and gallium phosphide metasurfaces
  exhibiting some of the largest nonlinear conversion efficiencies
  predicted thus far. Such enhancements arise because, in addition to
  enabling a great degree of tunability in the choice of design
  wavelengths, these optimization tools ensure both frequency- and
  phase-matching in addition to large nonlinear overlap factors,
  potentially orders of magnitude larger than those obtained in
  hand-designed structures.
\end{abstract}

\pacs{Valid PACS appear here}
\maketitle


\section{\label{sec:intro}Introduction}

Nonlinear frequency conversion plays a crucial role in many photonic
applications, including ultra-short pulse shaping~\cite{DeLong94,
  Arbore97}, spectroscopy~\cite{Heinz82}, generation of novel optical
states~\cite{Kuo06,Vodopyanov06,Krischek10}, and quantum information
processing~\cite{Vaziri02, Tanzilli05, Zaske12}.  Although frequency
conversion has been studied exhaustively in bulky optical systems,
including large ring-resonators~\cite{Furst10} and etalon
cavities~\cite{fejer1994nonlinear}, it remains largely unstudied in
micro- and nano-scale structures where light can be confined to
lengthscales on the order or even smaller than its wavelength. By
confining light over long times and small volumes, such highly compact
devices greatly enhance light--matter interactions, enabling similar
as well as new~\cite{soljavcic2004enhancement} functionalities
compared to those available in bulky systems but at much lower power
levels. Several proposals have been put forward based on the premise
of observing enhanced nonlinear effects in structures capable of
supporting multiple resonances at far-away
frequencies~\cite{Dumeige06,Wu87,Simonneau:97,Paschotta94,Koch99,Liscidini04,rivoire11:apl,Ramirez11,Zin14},
among which are micro-ring resonators~\cite{Pernice12,Bi12} and
photonic crystal cavities~\cite{Rivoire09,Buckley14}. However, to
date, these conventional designs fall short of simultaneously meeting
the many design challenges associated with resonant frequency
conversion, chief among them being the need to support multiple modes
with highly concentrated fields, exactly matched resonant frequencies,
and strong mode overlaps~\cite{Rodriguez07:OE}. Recently, we proposed
to leverage powerful, large-scale optimization techniques (commonly
known as inverse design) to allow computer-aided photonic designs that
can address all of these challenges.

\begin{figure}[htbp]
  \begin{center}
    \includegraphics[width=1\linewidth]{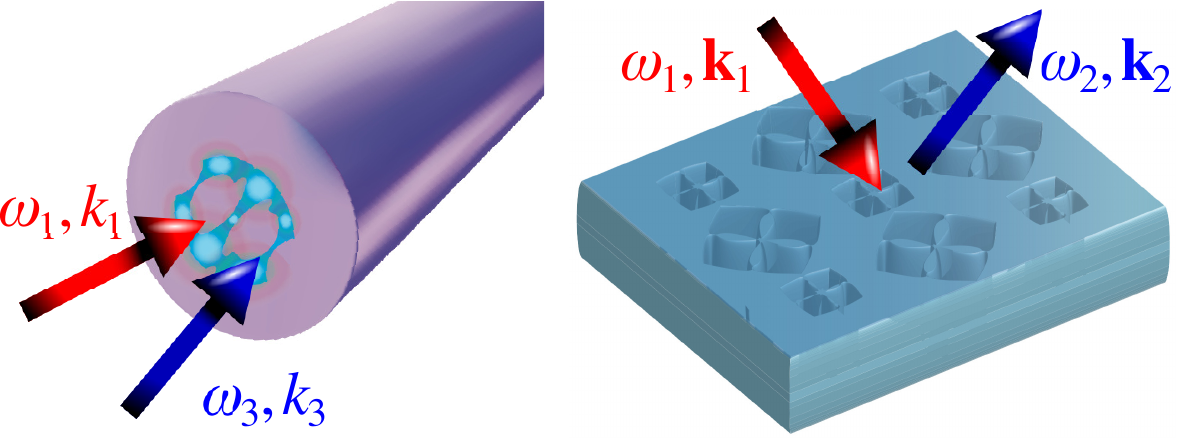}
    \caption{Schematic illustration of third-harmonic generation (THG)
      and second-harmonic generation (SHG) processes in
      inverse-designed microstructured fibers and metasurfaces,
      respectively.}
    \label{fig:scheme}
  \end{center}
\end{figure}

Our recently demonstrated optimization framework allows automatic
discovery of novel cavities supporting tightly localized modes at
several desired wavelengths and exhibiting large nonlinear mode
overlaps. As a proof-of-concept, we proposed doubly-resonant
structures, including multi-layered, aperiodic micro-post cavities and
multi-track ring resonators, capable of realizing second-harmonic
generation efficiencies exceeding
$10^4~\mathrm{W^{-1}}$~\cite{Lin:16,ZinOL}. In this paper, we extend
and apply this optimization approach to design extended structures,
including micro-structured optical fibers and photonic-crystal (PhC)
metasurfaces as shown in \figref{scheme}, for achieving
high-efficiency (second- and third-harmonic) frequency
conversion. Harmonic generation, which underlies numerous applications
in science, including coherent light sources~\cite{Lew95}, optical
imaging and microscopy~\cite{Yelin99,SHGMicroscpy} and
entangled-photon generation~\cite{Hamel10}, is now feasible at lower
power requirements thanks to the availability of highly nonlinear
$\chi^{(2)}$ and $\chi^{(3)}$ materials such as III-V semiconductor
compounds~\cite{rivoire11b:apl,Buckley13} and novel types of
chalcogenide glasses~\cite{Hall89}. In combination with advances in
materials synthesis, emerging fabrication technologies have also
enabled demonstrations of sophisticated micro-structured
fibers~\cite{ahmad2004high} and
metasurfaces~\cite{lapine2014colloquium,campione2014second,lee2014giant,wolf2015phased,o2015predicting,yang2015nonlinear,segal2015controlling,butet2015optical},
paving the way for experimental realization of inverse-designed
structures of increased geometric and fabrication complexity but which
offer orders-of-magnitude enhancements in conversion efficiencies and
the potential for augmented functionalities.

Given a material system of intrinsic $\chi^{(2)}$ or $\chi^{(3)}$
nonlinear coefficient, the efficiency of any given
frequency-conversion process in a resonant geometry will be determined
by a few modal parameters.  The possibility of confining light within
small mode volumes over long times or distances leads to significant
gains in efficiency (i.e. lower power requirements), stemming from the
higher intensity and cascadability of nonlinear interactions
(compensating for the otherwise small bulk nonlinearities). In
particular, the efficiency of such resonant processes depends on the
product of mode lifetimes and a nonlinear coefficient $\beta$, given
by \eqreftwo{b3}{b2} below, which generalizes the familiar concept of
quasi-phase matching to situations that include wavelength-scale
resonators~\cite{Rodriguez07:OE}. For propagating modes, leaky or
guided, the existence of a propagation phase further complicates this
figure of merit, with optimal designs requiring: (1) phase-matching
and frequency-matching conditions, (2) large nonlinear mode overlaps
$\beta$ and (3) large dimensionless lifetimes $Q$ (low material
absorption and/or radiative losses in the case of leaky modes). The
main design challenge is the difficult task of forming a doubly
resonant cavity with far-apart modes that \emph{simultaneously}
exhibit long lifetimes, large $\beta$, along with phase and frequency
matching. To date, the majority of prior works on frequency conversion
in fibers~\cite{Betourne08} and
metasurfaces~\cite{li2017nonlinear,wolf2015phased,lee2014giant,campione2014second,liu2016resonantly,wolf2015enhanced,yang2015nonlinear,tymchenko2015gradient},
have focused on only one of these aspects (usually phase-matching)
while ignoring the others. The geometries discovered by our
optimization framework, in contrast, address the above criteria,
revealing complex fibers and metasurfaces supporting TE or TM modes
with guaranteed phase and frequency matching, long lifetimes $Q$ and
enhanced overlap factors $\beta$ at any desired propagation
wavevector, and resulting in orders-of-magnitude enhancements in
conversion efficiencies.

\section{\label{sec:topo}An overview of optimization}

The possibility of fine-tuning spatial features of photonic devices to
realize functionalities not currently achievable by conventional
optical design methodologies based on index guiding and band-gap
confinement (which work exceedingly well but are otherwise limited for
narrowband applications), has been a major drive behind the last
several decades of interest in the topic of photonic
optimization~\cite{JoannopoulosJo08-book}. Among these techniques are
probabilistic Monte~Carlo algorithms, e.g. particle swarms, simulated
annealing, and genetic algorithms~\cite{Kim:04, Darki10,
  Minkov14}. Though sufficient for majority of narrow-band
(single-mode) applications, many of these gradient-free methods are
limited to typically small sets of design parameters~\cite{Chen06}
which often prove inadequate for handling wide-band (multi-mode)
problems. On the other hand, gradient-based inverse design techniques
are capable of efficiently exploring a much larger design space by
making use of analytical derivative information of the specified
objective and constraint functions~\cite{Jensen11}. Recently, the
development of versatile mathematical programming methods and the
rapid growth in computational power have enabled concurrent progress
in photonic inverse design, allowing theoretical (and more rencently,
experimental) demonstrations of complex topologies and unintuitive
geometries with unprecedented functionalities that would be arguably
difficult to realize through conventional intuition alone. However, to
date, most applications of inverse design in photonics are confined to
linear devices such as mode converters, waveguide bends and beam
splitters~\cite{Chen06,Jensen11,Liang13,Liu13,Piggott14,MenLee14,Piggott15,Shen15}. We
believe that this paper along with our recent
works~\cite{Lin:16,ZinOL} provide a glimpse of the potential of
photonic optimization in nonlinear optics.

A typical optimization problem seeks to maximize or minimize an
objective function $f$, subject to certain constraints $g$, over a set
of free variables or degrees of freedom
(DOF)~\cite{StrangComp}. Generally, one can classify photonic inverse
design into two different classes of optimization strategies, based
primarily on the nature or choice of
DOF~\cite{deaton2014survey}. Given a computational domain or grid, the
choice of a finite-dimensional parameter space not only determines the
degree of complexity but also the convergence and feasibility of the
solutions. One possibility is to exploit each DOF in the computational
domain as an optimization parameter, known as topology optimization
(TO), in which case one typically (though not always) chooses the
dielectric permittivity of each pixel $\epsilon(\mathbf{r})$ as a
degree of freedom (known as a continuous relaxation
parameter~\cite{bendsoe2004topology}). Another possibility, known as
shape optimization, is to expand the optimization parameter space in a
finite set of shapes (independent of the computational
discretization), which may be freeform contours represented by
so-called level sets~\cite{LvS} (the level-set method) or basic
geometric entities with simpler parametrizations
(e.g. polytopes)~\cite{haslinger2003introduction}. In the level-set
method, the zeros of a level-set ``function'' $\Phi(\mathbf{r})$
define the boundaries of ``binary shapes''; the optimization then
proceeds via a level-set PDE characterized by a velocity field which
is, in turn, constructed from derivative information~\cite{LvS}. A
much simpler variant (which we follow) is to choose a fixed but
sufficient number of basic binary shapes whose parameters can be made
to evolve by an optimization algorithm. Essentially, for such a
parametrization, the mathematical representations of the shapes must
yield continuous (analytic) derivatives, which is not feasible a
priori due to the finite computational discretization and can instead
be enforced by the use of a ``smoothing Kernel'' (described below).

A generic TO formulation is written down as:  
\begin{align}
  \text{max}/\text{min}\, &f(\epsbar_\alpha) \\
  &g(\epsbar_\alpha) \le 0 \\
  &0 \le \epsbar_\alpha \le 1
\end{align}
where the DOFs are the normalized dielectric permittivities
$\epsbar_\alpha \in [0, 1]$ assigned to each pixel or voxel (indexed
$\alpha$) in a specified volume~\cite{Jensen11,Liang13}. The subscript
$\alpha$ denotes appropriate spatial discretization
$\vecr \rightarrow (i,j,k)_\alpha \Delta$ with respect to Cartesian or
curvilinear coordinates. Depending on the choice of background (bg)
and structural materials, $\epsbar_\alpha$ is mapped onto
position-dependent dielectric constant via
$\e_\alpha = \lp \e - \e_\text{bg} \rp \epsbar_\alpha +
\e_\text{bg}$.
The binarity of the optimized structure is enforced by penalizing the
intermediate values $\epsbar \in (0,1)$ or utilizing a variety of
filter and regularization methods~\cite{Jensen11}. Starting from a
random initial guess, the technique discovers complex structures
automatically with the aid of powerful gradient-based algorithms such
as the method of moving asymptotes (MMA)~\cite{Svanberg02}. For an
electromagnetic problem, $f$ and $g$ are typically functions of the
electric $\vec{E}$ or magnetic $\vec{H}$ fields integrated over some
region, which are in turn solutions of Maxwell's equations under some
incident current or field. In what follows, we exploit direct solution
of Maxwell's equations,
\begin{align}
  \nabla \times {1 \over \mu}~\nabla \times \vec{E} -~
  \epsilon(\mathbf{r}) \omega^2 \vec{E} = i \omega \mathbf{J},
\label{eq:ME}
\end{align}
describing the steady-state field $\vec{E}(\vecr;\w)$ in response to
incident currents $\vec{J}(\vecr,\w)$ at frequency $\w$.  While
solution of \eqref{ME} is straightforward and commonplace, the key to
making optimization problems tractable is to obtain a fast-converging
and computationally efficient adjoint formulation of the
problem. Within the scope of TO, this requires efficient calculations
of the derivatives ${\partial f \over \partial
  \epsbar_\alpha},~{\partial g \over \partial \epsbar_\alpha}$ at
every pixel $\alpha$, which we perform by exploiting the
adjoint-variable method (AVM)~\cite{Jensen11}.

While the TO technique is quite efficient in handling the enormity of
an unconstrained design space, it often leads to geometries with
irregular features that are difficult to fabricate. An alternative
approach that is in principle more conducive to fabrication
constraints is to exploit shape optimization. In this work, we
primarily focus on a simple implementation of the latter that employs
a small and hence limited set of elementary geometric shapes,
e.g. ellipses~\cite{wang2017optimization} and polytopes, parametrized
by a few DOFs. In particular, we express the dielectric profile of the
computational domain as a sum of basic shape functions with
permittivities, $\epsbar_\alpha=\sum_\beta
H_\beta(\mathbf{r}_\alpha;\{p_\beta\})$, described by shape functions
$H_\beta$ and a finite set of geometric parameters $\{p_\beta\}$,
where $\beta$ denotes the shape index. Here, to deal with potential
overlap of two or more shapes, we implement a filter function that
enforces the same maximum-permittivity constraint $\bar{\epsilon} \leq
1$ described above. The derivatives of a given objective function $f$
(and associated constraints) can then be obtained via the chain rule
${\partial f \over \partial p_i} = {\partial f \over \partial
  \epsbar_\alpha} {\partial \epsbar_\alpha \over \partial p_i}$, where
the smoothness of the derivatives is guaranteed by insisting that the
shape functions $H$ be continuously differentiable functions. Below,
we choose non-piecewise-constant ellipsoidal shapes with exponentially
varying dielectric profiles near the boundaries, the smoothness of
which is determined by a few simple parameters that can, at various
points along the optimization, be slowly adjusted to realize fully
binary structures upon convergence. Such a ``relaxation''
process~\cite{haslinger2003introduction} is analogous to the
application of a binary filter in the objective
function~\cite{Jensen11}.

Any NFC process can be viewed as a frequency mixing scheme in which
two or more \emph{constituent} photons at a set of frequencies
$\{\w_n\}$ interact to produce an output photon at frequency
$\Omega=\sum_n c_n \w_n$, where $\{c_n\}$ can be either negative or
positive, depending on whether the corresponding photons are created
or destroyed in the process~\cite{Boyd92}. Given an appropriate
nonlinear tensor component $\chi_{ijk...}$, with
$i,j,k,...\in\{x,y,z\}$, mediating an interaction between the
polarization components $E_i(\Omega)$ and $E_{1j}$, $E_{2k}, ...$, we
begin with a collection of point dipole currents, each at the
\emph{constituent} frequency $\w_n,~n\in\{1,2,...\}$, such that
$\mathbf{J}_n = \hat{\mathbf{e}}_{n \nu}
\delta(\mathbf{r}-\mathbf{r}')$, where $\hat{\mathbf{e}}_{n \nu}
\in\{\hat{\mathbf{e}}_{1j},~\hat{\mathbf{e}}_{2k}, ...\}$ is a
polarization vector chosen so as to excite the desired electric-field
polarization components ($\nu$) of the corresponding mode at an
appropriate position $\mathbf{r}'$. Given the choice of incident
currents $\mathbf{J}_n$, we solve Maxwell's equations to obtain the
corresponding \emph{constituent} electric-field response $\vec{E}_n$,
from which one can construct a nonlinear polarization current
$\mathbf{J}(\Omega) = \bar{\epsilon}(\mathbf{r}) \prod_{n}
E_{n\nu}^{|c_n| (*)} \hat{\mathbf{e}}_i$, where $E_{n\nu}= \vE_n \cdot
\hat{\mathbf{e}}_{n\nu}$ and $\mathbf{J}(\Omega)$ can be generally
polarized ($\hat{\mathbf{e}}_i$) in a (chosen) direction that differs
from the constituent polarizations $\hat{\mathbf{e}}_{n\nu}$.  Here,
(*) denotes complex conjugation for negative $c_n$ and no conjugation
otherwise.  Finally, maximizing the radiated power, $-
\mathrm{Re}\Big[ \int \mathbf{J}(\Omega)^* \cdot \mathbf{E}(\Omega)
~d\mathbf{r} \Big]$, due to $\mathbf{J}(\Omega)$, one is immediately
led to the following nonlinear optimization problem:
\begin{align}
  \text{max}_{\bar{\epsilon}} ~ f(\bar{\epsilon};\w_n) &= -
  \mathrm{Re}\Big[ \int \mathbf{J}(\Omega)^* \cdot \mathbf{E}(\Omega)
    ~d\mathbf{r} \Big], \label{eq:ps1}\\ {\cal
    M}(\bar{\epsilon},\omega_n) \mathbf{E}_n &= i \omega_n
  \mathbf{J}_n,~ \mathbf{J}_n = \hat{\mathbf{e}}_{n \nu}
  \delta(\mathbf{r}-\mathbf{r}'), \notag \\ {\cal
    M}(\bar{\epsilon},\Omega) \mathbf{E}(\Omega) &= i \Omega
  \mathbf{J}(\Omega),~ \mathbf{J}(\Omega) = \bar{\epsilon} \prod_{n}
  E_{n \nu}^{|c_n| (*)} \hat{\mathbf{e}}_i, \notag \\ {\cal
    M}(\bar{\epsilon},\omega) &= \nabla \times {1 \over \mu}~\nabla
  \times -~ \epsilon(\mathbf{r}) \omega^2, \notag,
\end{align} 
where $\bar{\epsilon}$ is given by either the topology or shape
parameterizations described above. Writing down the objective function
in terms of the nonlinear polarization currents, it follows that
solution of \eqref{ps1}, obtained by employing any mathematical
programming technique that makes use of gradient information, e.g. the
adjoint variable method~\cite{Jensen11}, maximizes the nonlinear
coefficient (mode overlap) associated with the aforementioned
nonlinear optical process. The above framework can be easily extended
to consider propagating modes once we take into account the
appropriate Bloch boundary conditions that may arise from any desired
wave vectors imposed at the requisite frequencies~\cite{Taflove00}. In
the case of optical fibers or PhC metasurfaces (or, more generally,
any waveguiding system), such an extension naturally guarantees
perfect phase- and frequency-matching of the relevant modes in the
optimized structure.

\section{Third harmonic generation in fibers}

Conventional microstructured fibers (e.g. Bragg and holey fibers) are
typically designed based on intuitive principles like index-guiding
and bandgap confinement~\cite{JoannopoulosJo08-book}, and thus often
consist of periodic cross-sections comprising simple
shapes~\cite{Fink02,Feng03}. Below, we apply the aforementioned
optimization techniques to propose much more complicated
heterostructure fibers designed to enhance third-harmonic generation
at any desired wavelength. In order to achieve large third-harmonic
generation efficiencies, the fiber must support two co-propagating
modes of frequencies $\omega_1$ and $\omega_3=3\omega_1$ and
wavenumbers satisfying the phase-matching condition $k_3 =
3k_1$. Furthermore, the system must exhibit low radiative/dissipative
losses or alternatively, attenuation lengths that are much longer than
the corresponding interaction lengths $L$, defined as the propagation
length at which 50\% of the fundamental mode is up-converted. In the
small-input signal regime, the converted, third-harmonic output power
$P_3 \propto P_1^2$ and the interaction length $L = \frac{16}{3 k_1
  Z_0 |\beta_3| P_1}$ depend on the incident power $P_1$, vacuum
impedance $Z_0$ and nonlinear overlap factor~\cite{Grubsky:05}, \small
\begin{align}
\label{eq:b3}
  \beta_3 = \frac{\oiint_{S} \chi^{(3)} (\mathbf{E}_1^* \cdot
    \mathbf{E}_3) (\mathbf{E}_1^* \cdot \mathbf{E}_1^*)~dS}{
    \left(\textrm{Re}[\frac{1}{2}\oiint \sqrt{(\mathbf{E}_1^* \times
        \mathbf{H}_1)\cdot \hat{z}~dS]}\right)^{3}~
    \sqrt{\textrm{Re}[\frac{1}{2}\oiint (\mathbf{E}_3^* \times
      \mathbf{H}_3)\cdot \hat{z}~dS]}}
\end{align}
\normalsize which involves a complicated spatial overlap of the two
modes over the cross-sectional surface $S$ of the fiber. Note that the
attenuation coefficient $\gamma \equiv \omega/2 v_g Q$ of each mode
(the inverse of their respective attenuation length) is proportional
to their lifetime $Q$ and group velocity $v_g$.

\begin{figure}[t!]
  \begin{center}
    \includegraphics[width=0.9\linewidth]{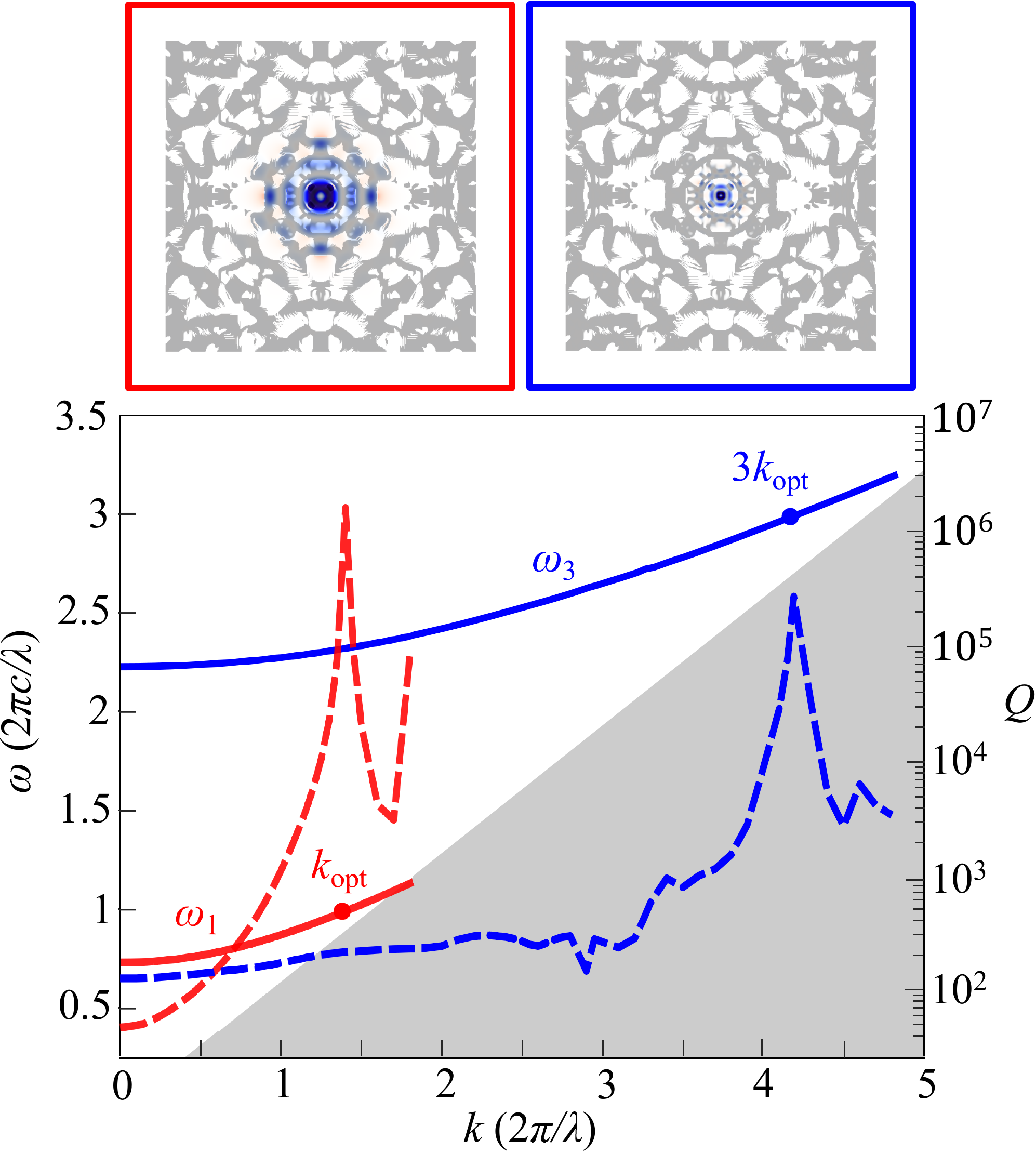}
    \caption[Dispersion relation, $Q\;\textrm{versus}\;k$, and mode
    profiles]{Dispersion relations and radiative lifetimes $Q$ versus
      propagation wavenumber $k$ of $\text{TM}_{01}$ fundamental
      $\omega_1$ (red) and third-harmonic $\omega_3$ (blue) modes in a
      chalcogenide/PES fiber optimized to achieve frequency matching
      $\omega_3=3\omega_1$ and large nonlinear overlaps at
      $k_\text{opt} = 1.4~(2\pi/\lambda)$. The shaded area in gray
      indicates regions laying below the chalcogenide light cone. The
      top insets show the fiber cross-section and coresponding power
      densities. }
    \label{fig:figure1}
  \end{center}
\end{figure}

\begin{figure}[t!]
  \begin{center}
    \includegraphics[width=1\linewidth]{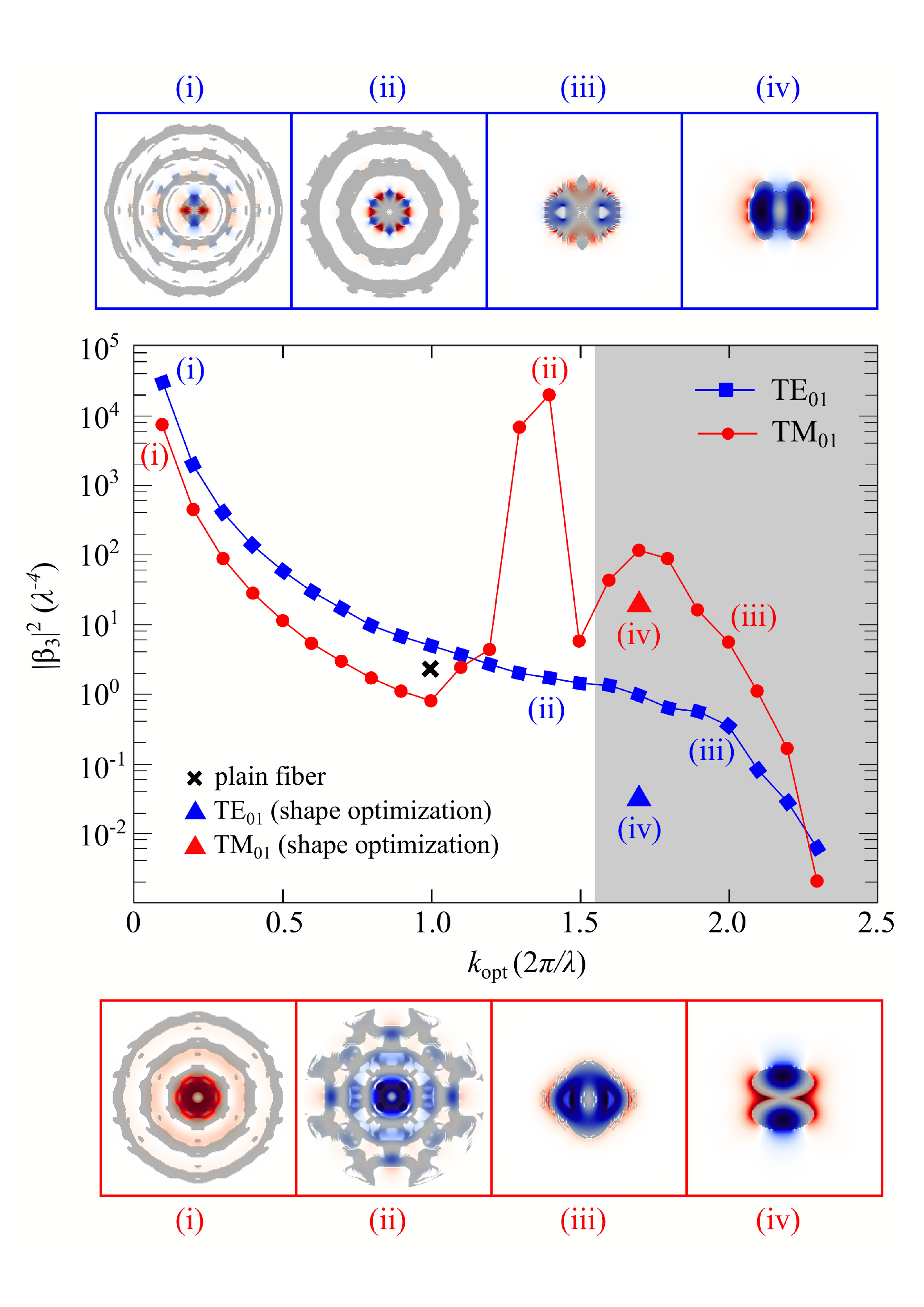}
    \caption[Optimal $|\beta_3|^2$ versus $k_\text{opt}$]{Nonlinear
      overlap factor $|\beta_3|^2$ corresponding to fundamental and
      third-harmonic modes in fibers which have been optimized to
      ensure phased-matched modes ($k_3=3k_\text{opt}$) at various
      fundamental-mode propagation wavenumbers $k_\text{opt}$, for
      both $\text{TE}_{01}$ (blue) and $\text{TM}_{01}$ (red)
      polarizations. The gray-shaded area denotes the regime of guided
      modes below the chalcogenide light line. For comparison, also
      shown is $|\beta_3|^2$ (black cross) of a standard plain fiber
      manually designed for operation at $\omega_1 = 0.914~(2\pi
      c/\lambda)$ and $k_1 =
      0.992~(2\pi/\lambda)$~\cite{Grubsky:05}. Shown as insets are
      fiber cross-sections along with power densities of fundamental
      modes at four different $k_\text{opt} = \{0.1, 1.4, 1.7,2.0\}
      (2\pi/\lambda)$.}
    \label{fig:figure2}
  \end{center}
\end{figure}

We focus on fibers comprising Chalcogenide/PES composites of
permittivities $\epsilon_\text{As$_2$Se$_3$} = 5.8125$ and
$\epsilon_\text{PES} = 2.4025$ at telecom wavelengths. Although our
technique can be readily applied to design the requisite properties at
any given wavenumber $k$ and for any desired polarization, we
specifically focus on designs for operation at wavenumbers in the
range $0.1~(2\pi/\lambda) < k_\text{opt} < 2.3~(2\pi/\lambda)$, with
$\lambda$ denoting the corresponding vacuum wavelength and
$k_\text{opt}$ the optimized wavenumber. We consider both leaky and
guided modes above and below the PES light line $\omega =
ck/\sqrt{\epsilon_\text{PES}}$, respectively, along with different
choices or transverse electric $\text{TE}_{01}$ and transverse
magnetic $\text{TM}_{01}$ polarizations. $\text{TE}_{01}$ modes are
those polarized along the plane of the fiber and consist primarily of
circulating $E_x$ and $E_y$ electric fields~\cite{fiberoptics}, while
$\text{TM}_{01}$ modes have electric fields $E_z$ polarized mainly
along the propagation direction $z$.

\Figref{figure1} (top insets) shows an inverse-designed fiber
cross-section that supports phased-matched $\text{TM}_{01}$
fundamental and third-harmonic modes (with profiles superimposed on
the insets) at $k_\text{opt} = k_1 = 1.4~(2\pi/\lambda)$. To ensure
that the optimization algorithm selectively finds $\text{TM}_{01}$
modes, we employ a magnetic current $\mathbf{J}_1 \sim \nabla\times
\delta(\mathbf{r})\hat{z}$ as the source in \eqref{ps1}, resulting in
electric fields of the desired polarization. The fiber cross-section
is represented by a $3\lambda \times 3\lambda$ computational cell
consisting of $300 \times 300$ pixels, where the size of each pixel is
$0.01\lambda \times 0.01\lambda$. From \Figref{figure1} (inset), it is
clear that both the fundamental and third harmonic modes are well
confined to the fiber core and exhibit substantial modal overlaps,
while again, the phase-matching condition is automatically satisfied
by the optimization process, with $k_3 = 3k_\text{opt}$. We find that
$|\beta_3|^2 \approx \num{2e4}~(\chi^{(3)}/\lambda^{4})$ is almost
four orders of magnitude larger than what has been demonstrated in
standard plain fibers, which have typical values of $|\beta_3|^2
\lesssim 2~(\chi^{(3)} /
\lambda^{4})$~\cite{Grubsky:05}. \Figref{figure1} shows the dispersion
of the two leaky modes (solid lines), with the PES lightline
represented by the gray region and their corresponding dimensionless
lifetimes, around $Q_1\approx 10^6$ and $Q_3\approx 10^5$ at
$k_\text{opt}$, plotted as dashed lines. Noticeably, while the fiber
is optimized to ensure phase matching at a single $k_\text{opt}$, any
phase mismatch remains small in the vicinity of $k \approx
k_\text{opt}$. In fact, even for $k \ll k_\text{opt}$, the frequency
difference is found to be only around $1\%$. Technically the only
factor limiting the lifetimes is the finite computational
cross-section (imposed by the finite computational cell), with much
larger lifetimes possible for larger cross sections. Away from
$k_\text{opt}$, the quality factors decrease while remaining
relatively large over a wide range of $k$. Considering the group
velocity $v_g$ around $k_\text{opt}$, we find that the attenuation
length of the fiber $L_\text{rad} = 1/\gamma \approx 2 v_g Q / \omega
= 1.66 \times 10^5 \lambda$. We note that while the fiber supports
mutliple modes around these wavelengths, the only modes near
$k_\text{opt}$ are those discovered by the optimization and shown in
the figure.

\Figref{figure2} shows the $\beta_3$ corresponding to fibers optimized
for operation at different values of $k_\text{opt}$ and polarizations,
and obtained by application of either topology (squares) or shape
(triangles) optimization.  The figure shows a general trend in which
$\beta_3$ decreases with increasing $k_\text{opt}$ for both
polarizations, except that $\text{TM}_{01}$ fibers tend to exhibit
non-monotonic behavior, with $\beta_3$ increasing sharply at an
intermediate $k_\text{opt} \approx (2\pi/\lambda)$ below the
lightline, above which it drops significantly before increasing again
in the guided regime, peaking again at $k_\text{opt} \approx
1.7~(2\pi/\lambda)$ before plummeting once again. We suspect that this
complicated behavior is not a consequence of any fundamental
limitation or physical consideration, but rather stems from the
optimization algorithm getting stuck in local minima. Regardless, our
results provide a proof-of-principle of the existence of fiber designs
with performance characteristics that can greatly surpass qthose of
traditional, hand-designed fibers. Furthermore, \figref{figure2} shows
typical fiber cross-sections at selective $k_\text{opt}$, along with
their corresponding superimposed (fundamental) mode profiles,
illustrating the fabricability of the structures.

\begin{table}[btp]
  \centering
  \caption{Representative second-harmonic generation figures of merit for both
    hand- and inverse-designed metasurfaces, including $\chi^{(2)}$, fundamental wavelength $\lambda_1$, and conversion efficiency $\eta$ per unit cell. Most of the literature only states the net conversion efficiency for metasurfaces with finite $N$ unit cells, which we convert to the efficiency $\eta$ per unit cell.}
  \begin{tabular}{cccc}
    \hline
    Structure &$\chi^{(2)}$ ($\text{nm}/V$) & $\lambda_1 (\mu \mathrm{m})$ & $\eta/(\chi^{(2)})^2$  \\ \hline
    
    gold split resonators~\cite{campione2014second} & 250 &10 &$2.1\times 10^{11}$\\

    gold split resonators~\cite{wolf2015enhanced} & 1.3 & 3.4 &$3.8\times 10^{11}$*\\
    
    gold cross bars~\cite{lee2014giant}&54&8&$1.4\times 10^{13}$*\\
    
    all-dielectric cylinders~\cite{liu2016resonantly} &0.2 & 1.02 & $1.6\times 10^{17}$*\\
    
    optimized design [\figref{meta}] & 0.1& 1.2 &$9.6\times 10^{24}$ \\\hline
  \end{tabular}
  \label{tab:shg}
\end{table}

Finally, we provide estimates of the power requirements associated
with these fiber designs.  We find that for a $\text{TM}_{01}$ fiber
operating at $k_\text{opt}=1.4~(2\pi/\lambda)$ and at a wavelength of
$\lambda=1\mu$m, conversion efficiencies of $50\%$ can be attained at
relatively small pump powers $P_1 \approx 1.7~\mathrm{mW}$ over a
fiber segment $L \approx 3~\text{cm}$, while the corresponding
(radiative) attenuation lengths are $\approx 17$cm. For comparison,
plain silica fibers~\cite{Grubsky:05} exhibit mode-overlap factors
$\beta_3 \approx 2~(\chi^{(3)}/\lambda^{2})$, leading to conversion
efficiencies on the order of $10^{-8}~\%$ for the same input power and
fiber length. Hence, the optimized structures achieve considerably
($\sim 10^9$ times) higher conversion efficiencies, an improvement
that is only partially due to the larger $\chi^{(3)}$ of ChG compared
to glass (approximately $440$ times larger). In particular, defining
the normalized interaction fiber length $L (\chi^{(3)})$, which
removes any source of material enhancement, we find that the optimized
fiber leads to a factor of $10^4$ enhancement. Similarly, we find that
a $\text{TE}_{01}$ fiber operating at
$k_\text{opt}=0.2~(2\pi/\lambda)$ results in a factor of $10^3$
enhancement compared to plain fibers.

\section{Second harmonic generation in metasurfaces}

\begin{figure}[b!]
  \begin{center}
    \includegraphics[width=1\linewidth]{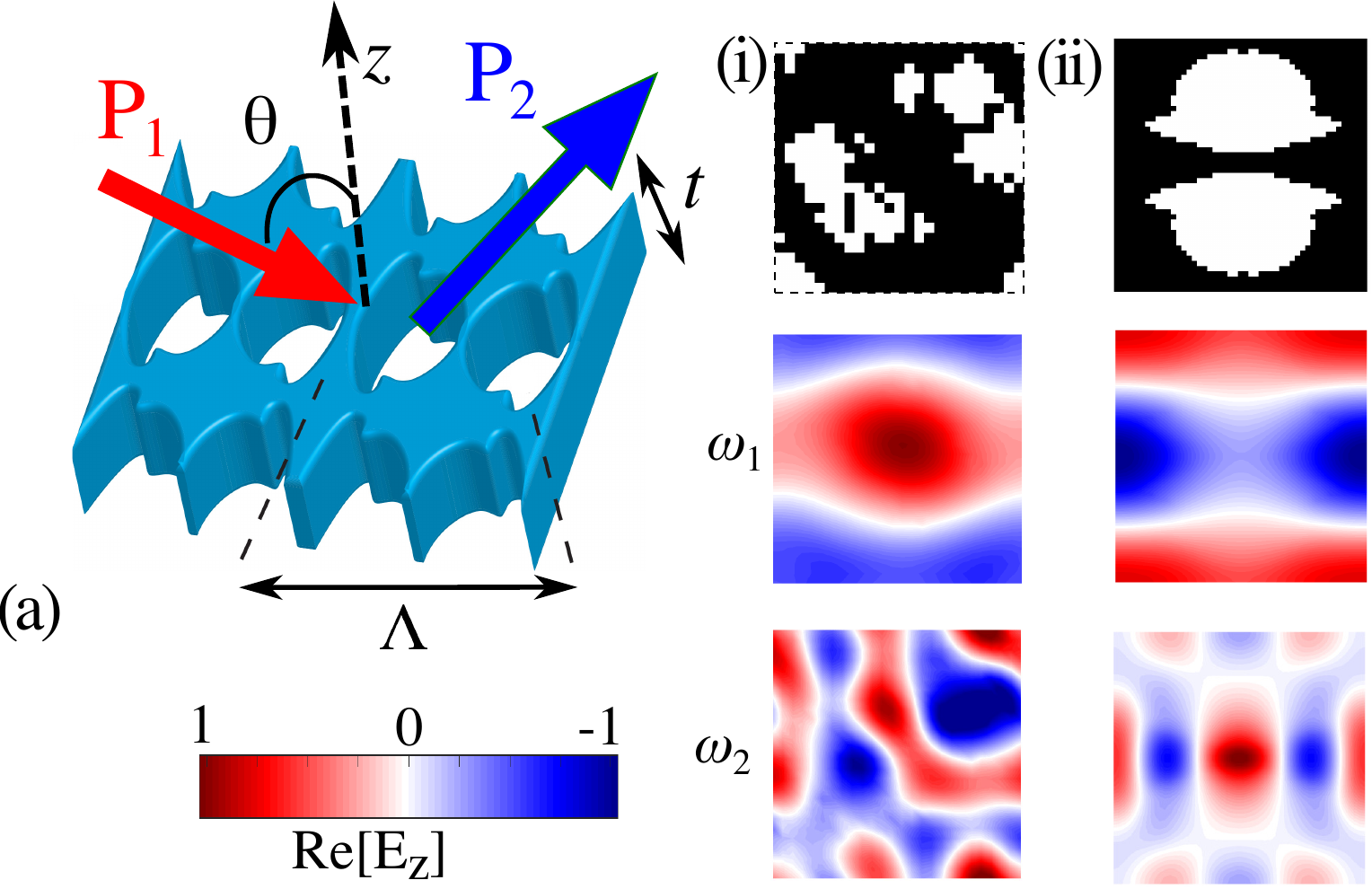}
    \caption{Schematic illustration of SHG in a square-lattice
      metasurface of finite thickness $t$ and period
      $\Lambda\times\Lambda$. Shown are dielectric profiles and mode
      profiles ($E_z$) over one unit cell for a single ($z=0$) cross
      section for two inverse-designed metasurfaces, optimized to
      ensure frequency- and phase-matching for light incident at at an
      angle $\theta=3^\circ$ (i) or normal incidence (ii) with respect
      to one of the axes. Dark (white) represents gallium phosphide
      (vacuum) regions.}
    \label{fig:meta}
  \end{center}
\end{figure}

Metasurfaces offer an advantageous platform for realizing complicated
beam generation and wavefront shaping over extended
surfaces~\cite{yu2014flat} and have recently been exploited in
conjunction with nonlinear materials as a means of generating and
controlling light at multiple
wavelengths~\cite{li2017nonlinear,michaeli2017nonlinear,keren2015nonlinear,segal2015controlling}. A
typical nonlinear metasurfaces can suffer from poor
frequency-conversion efficiencies due to a combination of weak
confinement, material absorption, and sub-optimal mode overlaps. In
particular, typical designs exploit
plasmonic~\cite{campione2014second,wolf2015enhanced,lee2014giant,wolf2015phased}
or all-dielectric~\cite{liu2016resonantly,yang2015nonlinear} elements
comprising simple shapes distributed over a unit cell, including split
ring
resonators~\cite{campione2014second,wolf2015phased,wolf2015enhanced},
cross-bars~\cite{lee2014giant}, and cylindrical
posts~\cite{liu2016resonantly}, with the main focus being that of
satisfying the requisite frequency- and phase-matching
condition~\cite{krasnok2017nonlinear}. Here, we show that inverse
design can not only facilitate the enforcement of frequency- and
phase-matching requirements but also allow further enhancements
stemming from the intentional engineering of nonlinear modal overlaps,
often neglected in typical designs.

To achive large second-harmonic generation efficiencies, a metasurface
must support two extended resonances at frequencies $\omega_1$ and
$\omega_2 = 2\omega_1$ and wavevectors satisfying the phase-matching
condition $\mathbf{k}_2=2\mathbf{k}_1$. As illustrated schematically
in \figref{meta}, a typical setup consists of an incident wave of
power per unit cell $P_1$ at some frequency and angle (described by
wavenumber $\mathbf{k}_1$) and a corresponding output harmonic wave of
power per unit cell, $P_2$. In the small-signal regime, the output
power $P_2 \propto P_1^2$ scales quadratically with $P_1$, resulting
in a conversion efficiency per unit cell, 
\begin{equation}
  \eta=\frac{P_2}{P_1^2} = \frac{Q_1^4Q_2^2}{Q_{1,\text{rad}}^2Q_{2,\text{rad}}}\frac{|\beta_2|^2\lambda_1}{\pi\varepsilon_0c}
\end{equation}
where $Q$ and $Q_\text{rad}$ denote total and radiative dimensionless
lifetimes and $\beta_2$ the nonlinear overlap factor,
\begin{equation}
\label{eq:b2}
  \beta_2=\frac{\int_{V} dV\, \chi^{(2)} \mathbf{E}_2^{*}\cdot \mathbf{E}_1^2}{\left( \int dV\,\epsilon_1|\mathbf{E}_1|^2 \right)\left( \sqrt{\int dV\, \epsilon_2|\mathbf{E}_2|^2} \right)}
\end{equation}
Note that the conversion efficiency is defined as the efficiency per
unit cell, since we are dealing with an extended surface.


We now apply our optimization framework to discover new all-dielectric
metasurfaces, with the permittivity of the medium
$\epsilon_\text{GaP}$ taken to be that of gallium phospide near
telecom
wavelengths~\cite{bond1965measurement,shoji1997absolute}. Note,
however, that the same framework can be easily extended to design
plasmonic surfaces.  The metasurfaces, illustrated schematically in
\figref{meta}, are square photonic-crystal slabs of in-plane
periodcity $\Lambda \times \Lambda$ and finite thickness $t$. To
ensure fabricability, the optimization parameters are taken to lie in
the plane of the metasurface, resulting in structures that can be
fabricated by etching.

\Figref{meta} shows cross-sections of the unit cell of two GaP
metasurfaces of thicknesses $t=612$~nm and $\Lambda=480$~nm, designed
for operation at a fundamental frequency $\omega_1=1.57\times
10^{15}$~rad/s ($\lambda=1.2$~$\mu$m) so as to satisfy both frequency-
and phase-matching conditions. Also shown are the corresponding
fundamental and harmonic mode profiles. The structure on the left is
optimized for operation at an incident angle $\theta\approx 3.6^\circ$
relative to the out-of-plane axis and is found to exhibit large
radiative lifetimes $Q_{1(2)}^\text{rad} \approx 6 (2) \times 10^4$
and overlap factor $|\beta_2|^2=1.6 \times 10^{-3}
(\chi^{(2)}/\lambda^3)$. The structure on the right is instead
optimized for operation at normal incidence, resulting in a slightly
smaller $|\beta_2|^2 = 4 \times 10^{-4} (\chi^{(2)}/\lambda^3)$. Due
to the symmetry of the structure, the modes exhibit infinite lifetimes
(and hence are technically dark modes), though in practice,
fabrication imperfections necessarily lead to finite lifetimes.
\Tabref{shg} compares a few of the relevant figures of merits for
representative metasurface designs, which include both plasmonic and
dielectric structures. Although comparing $\beta_2$ appears to be
impossible due to a surprising lack of relevant modal parameters in
these
studies~\cite{campione2014second,wolf2015enhanced,lee2014giant,liu2016resonantly}.
including the absence of radiative and dissipative quality factors, we
find that the optimized designs exhibit orders of magnitude larger
conversion efficiencies. While it is difficult to distinguish the
relative impact of the mode lifetimes and overlap factors, arguably,
the optimized structures overcome several limitations associated with
previous designs. On the one hand, plasmonic structures exhibit
tightly confined modes and therefore lead to large nonlinear overlaps,
but absorptive losses and weak material nonlinearities imply that they
suffer from small lifetimes. On the other hand, several of the
proposed all-dielectric metasurfaces have had negligible material
losses and hence larger lifetimes, but have not been designed to
ensure large nonlinear overlaps. 



\section{\label{sec:conclusion}Concluding Remarks}

We have demonstrated a novel optimization approach for the design of
nonlinear photonic fibers and metasurfaces. The optimized structures
demonstrate very high leaky-mode lifetimes for both fundamental and
harmonic modes and order-of-magnitude larger overlap factors than
traditional designs. Inverse design not only overcomes efficiency
limitations of traditional index fibers and photonic crystal
metasurfaces but also greatly reduces challenges and difficulties
inherent to the design process. Although in this report we have not
considered effects resulting from self- or cross-phase modulation, we
expect no significant impact on the conversion efficiency in the
small-signal limit since the finite bandwidth around the designated
phase-matched propagation wavevectors can potentially compensate for
any small phase-mismatch that might arise.  At larger powers where
these effects cannot be ignored, one could account and compensate for
them through minor modifications to the optimization objective
function, the subject of future work. Furthermore, we will consider
extending our inverse design framework to terahertz frequency
generation and other novel nonlinear processes.

\section{Acknowledgements}

This material is based upon work supported by the National Science
Foundation under Grant No. DMR-1454836, a MRSEC supported by NSF Grant
No. DMR 1420541, and NSF Award EFMA-1640986.


\begin{thebibliography}{80}%
\makeatletter
\providecommand \@ifxundefined [1]{%
 \@ifx{#1\undefined}
}%
\providecommand \@ifnum [1]{%
 \ifnum #1\expandafter \@firstoftwo
 \else \expandafter \@secondoftwo
 \fi
}%
\providecommand \@ifx [1]{%
 \ifx #1\expandafter \@firstoftwo
 \else \expandafter \@secondoftwo
 \fi
}%
\providecommand \natexlab [1]{#1}%
\providecommand \enquote  [1]{``#1''}%
\providecommand \bibnamefont  [1]{#1}%
\providecommand \bibfnamefont [1]{#1}%
\providecommand \citenamefont [1]{#1}%
\providecommand \href@noop [0]{\@secondoftwo}%
\providecommand \href [0]{\begingroup \@sanitize@url \@href}%
\providecommand \@href[1]{\@@startlink{#1}\@@href}%
\providecommand \@@href[1]{\endgroup#1\@@endlink}%
\providecommand \@sanitize@url [0]{\catcode `\\12\catcode `\$12\catcode
  `\&12\catcode `\#12\catcode `\^12\catcode `\_12\catcode `\%12\relax}%
\providecommand \@@startlink[1]{}%
\providecommand \@@endlink[0]{}%
\providecommand \url  [0]{\begingroup\@sanitize@url \@url }%
\providecommand \@url [1]{\endgroup\@href {#1}{\urlprefix }}%
\providecommand \urlprefix  [0]{URL }%
\providecommand \Eprint [0]{\href }%
\providecommand \doibase [0]{http://dx.doi.org/}%
\providecommand \selectlanguage [0]{\@gobble}%
\providecommand \bibinfo  [0]{\@secondoftwo}%
\providecommand \bibfield  [0]{\@secondoftwo}%
\providecommand \translation [1]{[#1]}%
\providecommand \BibitemOpen [0]{}%
\providecommand \bibitemStop [0]{}%
\providecommand \bibitemNoStop [0]{.\EOS\space}%
\providecommand \EOS [0]{\spacefactor3000\relax}%
\providecommand \BibitemShut  [1]{\csname bibitem#1\endcsname}%
\let\auto@bib@innerbib\@empty
\bibitem [{\citenamefont {DeLong}\ \emph {et~al.}(1994)\citenamefont {DeLong},
  \citenamefont {Trebino}, \citenamefont {Hunter},\ and\ \citenamefont
  {White}}]{DeLong94}%
  \BibitemOpen
  \bibfield  {author} {\bibinfo {author} {\bibfnamefont {K.~W.}\ \bibnamefont
  {DeLong}}, \bibinfo {author} {\bibfnamefont {R.}~\bibnamefont {Trebino}},
  \bibinfo {author} {\bibfnamefont {J.}~\bibnamefont {Hunter}}, \ and\ \bibinfo
  {author} {\bibfnamefont {W.~E.}\ \bibnamefont {White}},\ }\href@noop {}
  {\bibfield  {journal} {\bibinfo  {journal} {J.~Opt. Soc. Am.~B}\ }\textbf
  {\bibinfo {volume} {11}},\ \bibinfo {pages} {2206} (\bibinfo {year}
  {1994})}\BibitemShut {NoStop}%
\bibitem [{\citenamefont {Arbore}\ \emph {et~al.}(1997)\citenamefont {Arbore},
  \citenamefont {Galvanauskas}, \citenamefont {Harter}, \citenamefont {Chou},\
  and\ \citenamefont {Fejer}}]{Arbore97}%
  \BibitemOpen
  \bibfield  {author} {\bibinfo {author} {\bibfnamefont {M.~A.}\ \bibnamefont
  {Arbore}}, \bibinfo {author} {\bibfnamefont {A.}~\bibnamefont
  {Galvanauskas}}, \bibinfo {author} {\bibfnamefont {D.}~\bibnamefont
  {Harter}}, \bibinfo {author} {\bibfnamefont {M.~H.}\ \bibnamefont {Chou}}, \
  and\ \bibinfo {author} {\bibfnamefont {M.~M.}\ \bibnamefont {Fejer}},\
  }\href@noop {} {\bibfield  {journal} {\bibinfo  {journal} {Opt. Lett.}\
  }\textbf {\bibinfo {volume} {22}},\ \bibinfo {pages} {1341} (\bibinfo {year}
  {1997})}\BibitemShut {NoStop}%
\bibitem [{\citenamefont {Heinz}\ \emph {et~al.}(1982)\citenamefont {Heinz},
  \citenamefont {Chen}, \citenamefont {Ricard},\ and\ \citenamefont
  {Shen}}]{Heinz82}%
  \BibitemOpen
  \bibfield  {author} {\bibinfo {author} {\bibfnamefont {T.~F.}\ \bibnamefont
  {Heinz}}, \bibinfo {author} {\bibfnamefont {C.~K.}\ \bibnamefont {Chen}},
  \bibinfo {author} {\bibfnamefont {D.}~\bibnamefont {Ricard}}, \ and\ \bibinfo
  {author} {\bibfnamefont {Y.~R.}\ \bibnamefont {Shen}},\ }\href@noop {}
  {\bibfield  {journal} {\bibinfo  {journal} {Phys. Rev. Lett.}\ }\textbf
  {\bibinfo {volume} {48}},\ \bibinfo {pages} {478} (\bibinfo {year}
  {1982})}\BibitemShut {NoStop}%
\bibitem [{\citenamefont {Kuo}\ \emph {et~al.}(2006)\citenamefont {Kuo},
  \citenamefont {Vodopyanov}, \citenamefont {Fejer}, \citenamefont
  {Simanovskii}, \citenamefont {Yu}, \citenamefont {Harris}, \citenamefont
  {Bliss},\ and\ \citenamefont {D.Weyburne}}]{Kuo06}%
  \BibitemOpen
  \bibfield  {author} {\bibinfo {author} {\bibfnamefont {P.~S.}\ \bibnamefont
  {Kuo}}, \bibinfo {author} {\bibfnamefont {K.~L.}\ \bibnamefont {Vodopyanov}},
  \bibinfo {author} {\bibfnamefont {M.~M.}\ \bibnamefont {Fejer}}, \bibinfo
  {author} {\bibfnamefont {D.~M.}\ \bibnamefont {Simanovskii}}, \bibinfo
  {author} {\bibfnamefont {X.}~\bibnamefont {Yu}}, \bibinfo {author}
  {\bibfnamefont {J.~S.}\ \bibnamefont {Harris}}, \bibinfo {author}
  {\bibfnamefont {D.}~\bibnamefont {Bliss}}, \ and\ \bibinfo {author}
  {\bibnamefont {D.Weyburne}},\ }\href@noop {} {\bibfield  {journal} {\bibinfo
  {journal} {Opt. Lett.}\ }\textbf {\bibinfo {volume} {31}},\ \bibinfo {pages}
  {71} (\bibinfo {year} {2006})}\BibitemShut {NoStop}%
\bibitem [{\citenamefont {Vodopyanov}\ \emph {et~al.}(2006)\citenamefont
  {Vodopyanov}, \citenamefont {Fejer}, \citenamefont {Yu}, \citenamefont
  {Harris}, \citenamefont {Lee}, \citenamefont {Hurlbut}, \citenamefont
  {Kozlov}, \citenamefont {Bliss},\ and\ \citenamefont {Lynch}}]{Vodopyanov06}%
  \BibitemOpen
  \bibfield  {author} {\bibinfo {author} {\bibfnamefont {K.~L.}\ \bibnamefont
  {Vodopyanov}}, \bibinfo {author} {\bibfnamefont {M.~M.}\ \bibnamefont
  {Fejer}}, \bibinfo {author} {\bibfnamefont {X.}~\bibnamefont {Yu}}, \bibinfo
  {author} {\bibfnamefont {J.~S.}\ \bibnamefont {Harris}}, \bibinfo {author}
  {\bibfnamefont {Y.-S.}\ \bibnamefont {Lee}}, \bibinfo {author} {\bibfnamefont
  {W.~C.}\ \bibnamefont {Hurlbut}}, \bibinfo {author} {\bibfnamefont {V.~G.}\
  \bibnamefont {Kozlov}}, \bibinfo {author} {\bibfnamefont {D.}~\bibnamefont
  {Bliss}}, \ and\ \bibinfo {author} {\bibfnamefont {C.}~\bibnamefont
  {Lynch}},\ }\href@noop {} {\bibfield  {journal} {\bibinfo  {journal} {Appl.
  Phys. Lett.}\ }\textbf {\bibinfo {volume} {89}},\ \bibinfo {pages} {141119}
  (\bibinfo {year} {2006})}\BibitemShut {NoStop}%
\bibitem [{\citenamefont {Krischek}\ \emph {et~al.}(2010)\citenamefont
  {Krischek}, \citenamefont {Wieczorek}, \citenamefont {Ozawa}, \citenamefont
  {Kiesel}, \citenamefont {Michelberger}, \citenamefont {Udem},\ and\
  \citenamefont {Weinfurter}}]{Krischek10}%
  \BibitemOpen
  \bibfield  {author} {\bibinfo {author} {\bibfnamefont {R.}~\bibnamefont
  {Krischek}}, \bibinfo {author} {\bibfnamefont {W.}~\bibnamefont {Wieczorek}},
  \bibinfo {author} {\bibfnamefont {A.}~\bibnamefont {Ozawa}}, \bibinfo
  {author} {\bibfnamefont {N.}~\bibnamefont {Kiesel}}, \bibinfo {author}
  {\bibfnamefont {P.}~\bibnamefont {Michelberger}}, \bibinfo {author}
  {\bibfnamefont {T.}~\bibnamefont {Udem}}, \ and\ \bibinfo {author}
  {\bibfnamefont {H.}~\bibnamefont {Weinfurter}},\ }\href@noop {} {\bibfield
  {journal} {\bibinfo  {journal} {Nature Photonics}\ }\textbf {\bibinfo
  {volume} {4}},\ \bibinfo {pages} {170} (\bibinfo {year} {2010})}\BibitemShut
  {NoStop}%
\bibitem [{\citenamefont {Vaziri}\ \emph {et~al.}(2002)\citenamefont {Vaziri},
  \citenamefont {Weihs},\ and\ \citenamefont {Zeilinger}}]{Vaziri02}%
  \BibitemOpen
  \bibfield  {author} {\bibinfo {author} {\bibfnamefont {A.}~\bibnamefont
  {Vaziri}}, \bibinfo {author} {\bibfnamefont {G.}~\bibnamefont {Weihs}}, \
  and\ \bibinfo {author} {\bibfnamefont {A.}~\bibnamefont {Zeilinger}},\ }\href
  {\doibase 10.1103/PhysRevLett.89.240401} {\bibfield  {journal} {\bibinfo
  {journal} {Phys. Rev. Lett.}\ }\textbf {\bibinfo {volume} {89}},\ \bibinfo
  {pages} {240401} (\bibinfo {year} {2002})}\BibitemShut {NoStop}%
\bibitem [{\citenamefont {Tanzilli}\ \emph {et~al.}(2005)\citenamefont
  {Tanzilli}, \citenamefont {Tittel}, \citenamefont {Halder}, \citenamefont
  {Alibart}, \citenamefont {Baldi}, \citenamefont {Gisin},\ and\ \citenamefont
  {Zbinden}}]{Tanzilli05}%
  \BibitemOpen
  \bibfield  {author} {\bibinfo {author} {\bibfnamefont {S.}~\bibnamefont
  {Tanzilli}}, \bibinfo {author} {\bibfnamefont {W.}~\bibnamefont {Tittel}},
  \bibinfo {author} {\bibfnamefont {M.}~\bibnamefont {Halder}}, \bibinfo
  {author} {\bibfnamefont {O.}~\bibnamefont {Alibart}}, \bibinfo {author}
  {\bibfnamefont {P.}~\bibnamefont {Baldi}}, \bibinfo {author} {\bibfnamefont
  {N.}~\bibnamefont {Gisin}}, \ and\ \bibinfo {author} {\bibfnamefont
  {H.}~\bibnamefont {Zbinden}},\ }\href@noop {} {\bibfield  {journal} {\bibinfo
   {journal} {Nature}\ }\textbf {\bibinfo {volume} {437}},\ \bibinfo {pages}
  {116} (\bibinfo {year} {2005})}\BibitemShut {NoStop}%
\bibitem [{\citenamefont {Zaske}\ \emph {et~al.}(2012)\citenamefont {Zaske},
  \citenamefont {Lenhard}, \citenamefont {Ke\ss{}ler}, \citenamefont {Kettler},
  \citenamefont {Hepp}, \citenamefont {Arend}, \citenamefont {Albrecht},
  \citenamefont {Schulz}, \citenamefont {Jetter}, \citenamefont {Michler},\
  and\ \citenamefont {Becher}}]{Zaske12}%
  \BibitemOpen
  \bibfield  {author} {\bibinfo {author} {\bibfnamefont {S.}~\bibnamefont
  {Zaske}}, \bibinfo {author} {\bibfnamefont {A.}~\bibnamefont {Lenhard}},
  \bibinfo {author} {\bibfnamefont {C.~A.}\ \bibnamefont {Ke\ss{}ler}},
  \bibinfo {author} {\bibfnamefont {J.}~\bibnamefont {Kettler}}, \bibinfo
  {author} {\bibfnamefont {C.}~\bibnamefont {Hepp}}, \bibinfo {author}
  {\bibfnamefont {C.}~\bibnamefont {Arend}}, \bibinfo {author} {\bibfnamefont
  {R.}~\bibnamefont {Albrecht}}, \bibinfo {author} {\bibfnamefont {W.-M.}\
  \bibnamefont {Schulz}}, \bibinfo {author} {\bibfnamefont {M.}~\bibnamefont
  {Jetter}}, \bibinfo {author} {\bibfnamefont {P.}~\bibnamefont {Michler}}, \
  and\ \bibinfo {author} {\bibfnamefont {C.}~\bibnamefont {Becher}},\ }\href
  {\doibase 10.1103/PhysRevLett.109.147404} {\bibfield  {journal} {\bibinfo
  {journal} {Phys. Rev. Lett.}\ }\textbf {\bibinfo {volume} {109}},\ \bibinfo
  {pages} {147404} (\bibinfo {year} {2012})}\BibitemShut {NoStop}%
\bibitem [{\citenamefont {F\"urst}\ \emph {et~al.}(2010)\citenamefont
  {F\"urst}, \citenamefont {Strekalov}, \citenamefont {Elser}, \citenamefont
  {Lassen}, \citenamefont {Andersen}, \citenamefont {Marquardt},\ and\
  \citenamefont {Leuchs}}]{Furst10}%
  \BibitemOpen
  \bibfield  {author} {\bibinfo {author} {\bibfnamefont {J.~U.}\ \bibnamefont
  {F\"urst}}, \bibinfo {author} {\bibfnamefont {D.~V.}\ \bibnamefont
  {Strekalov}}, \bibinfo {author} {\bibfnamefont {D.}~\bibnamefont {Elser}},
  \bibinfo {author} {\bibfnamefont {M.}~\bibnamefont {Lassen}}, \bibinfo
  {author} {\bibfnamefont {U.~L.}\ \bibnamefont {Andersen}}, \bibinfo {author}
  {\bibfnamefont {C.}~\bibnamefont {Marquardt}}, \ and\ \bibinfo {author}
  {\bibfnamefont {G.}~\bibnamefont {Leuchs}},\ }\href {\doibase
  10.1103/PhysRevLett.104.153901} {\bibfield  {journal} {\bibinfo  {journal}
  {Phys. Rev. Lett.}\ }\textbf {\bibinfo {volume} {104}},\ \bibinfo {pages}
  {153901} (\bibinfo {year} {2010})}\BibitemShut {NoStop}%
\bibitem [{\citenamefont {Fejer}(1994)}]{fejer1994nonlinear}%
  \BibitemOpen
  \bibfield  {author} {\bibinfo {author} {\bibfnamefont {M.~M.}\ \bibnamefont
  {Fejer}},\ }\href@noop {} {\bibfield  {journal} {\bibinfo  {journal} {Physics
  today}\ }\textbf {\bibinfo {volume} {47}},\ \bibinfo {pages} {25} (\bibinfo
  {year} {1994})}\BibitemShut {NoStop}%
\bibitem [{\citenamefont {Solja{\v{c}}i{\'c}}\ and\ \citenamefont
  {Joannopoulos}(2004)}]{soljavcic2004enhancement}%
  \BibitemOpen
  \bibfield  {author} {\bibinfo {author} {\bibfnamefont {M.}~\bibnamefont
  {Solja{\v{c}}i{\'c}}}\ and\ \bibinfo {author} {\bibfnamefont {J.~D.}\
  \bibnamefont {Joannopoulos}},\ }\href@noop {} {\bibfield  {journal} {\bibinfo
   {journal} {Nature materials}\ }\textbf {\bibinfo {volume} {3}},\ \bibinfo
  {pages} {211} (\bibinfo {year} {2004})}\BibitemShut {NoStop}%
\bibitem [{\citenamefont {Dumeige}\ and\ \citenamefont
  {Feron}(2006)}]{Dumeige06}%
  \BibitemOpen
  \bibfield  {author} {\bibinfo {author} {\bibfnamefont {Y.}~\bibnamefont
  {Dumeige}}\ and\ \bibinfo {author} {\bibfnamefont {P.}~\bibnamefont
  {Feron}},\ }\href@noop {} {\bibfield  {journal} {\bibinfo  {journal} {Phys.
  Rev.~A}\ }\textbf {\bibinfo {volume} {74}},\ \bibinfo {pages} {063804}
  (\bibinfo {year} {2006})}\BibitemShut {NoStop}%
\bibitem [{\citenamefont {Wu}\ \emph {et~al.}(1987)\citenamefont {Wu},
  \citenamefont {Xiao},\ and\ \citenamefont {Kimble}}]{Wu87}%
  \BibitemOpen
  \bibfield  {author} {\bibinfo {author} {\bibfnamefont {L.-A.}\ \bibnamefont
  {Wu}}, \bibinfo {author} {\bibfnamefont {M.}~\bibnamefont {Xiao}}, \ and\
  \bibinfo {author} {\bibfnamefont {H.~J.}\ \bibnamefont {Kimble}},\
  }\href@noop {} {\bibfield  {journal} {\bibinfo  {journal} {JOSA-B}\ }\textbf
  {\bibinfo {volume} {4}},\ \bibinfo {pages} {1465} (\bibinfo {year}
  {1987})}\BibitemShut {NoStop}%
\bibitem [{\citenamefont {Simonneau}\ \emph {et~al.}(1997)\citenamefont
  {Simonneau}, \citenamefont {Debray}, \citenamefont {Harmand}, \citenamefont
  {Vidakovi\'{c}}, \citenamefont {Lovering},\ and\ \citenamefont
  {Levenson}}]{Simonneau:97}%
  \BibitemOpen
  \bibfield  {author} {\bibinfo {author} {\bibfnamefont {C.}~\bibnamefont
  {Simonneau}}, \bibinfo {author} {\bibfnamefont {J.~P.}\ \bibnamefont
  {Debray}}, \bibinfo {author} {\bibfnamefont {J.~C.}\ \bibnamefont {Harmand}},
  \bibinfo {author} {\bibfnamefont {P.}~\bibnamefont {Vidakovi\'{c}}}, \bibinfo
  {author} {\bibfnamefont {D.~J.}\ \bibnamefont {Lovering}}, \ and\ \bibinfo
  {author} {\bibfnamefont {J.~A.}\ \bibnamefont {Levenson}},\ }\href@noop {}
  {\bibfield  {journal} {\bibinfo  {journal} {Opt. Lett.}\ }\textbf {\bibinfo
  {volume} {22}},\ \bibinfo {pages} {1775} (\bibinfo {year}
  {1997})}\BibitemShut {NoStop}%
\bibitem [{\citenamefont {Paschotta}\ \emph {et~al.}(1994)\citenamefont
  {Paschotta}, \citenamefont {Fiedler}, \citenamefont {Kurz},\ and\
  \citenamefont {Mlynek}}]{Paschotta94}%
  \BibitemOpen
  \bibfield  {author} {\bibinfo {author} {\bibfnamefont {R.}~\bibnamefont
  {Paschotta}}, \bibinfo {author} {\bibfnamefont {K.}~\bibnamefont {Fiedler}},
  \bibinfo {author} {\bibfnamefont {P.}~\bibnamefont {Kurz}}, \ and\ \bibinfo
  {author} {\bibfnamefont {J.}~\bibnamefont {Mlynek}},\ }\href@noop {}
  {\bibfield  {journal} {\bibinfo  {journal} {Appl. Phys. Lett.}\ }\textbf
  {\bibinfo {volume} {58}},\ \bibinfo {pages} {117} (\bibinfo {year}
  {1994})}\BibitemShut {NoStop}%
\bibitem [{\citenamefont {Koch}\ and\ \citenamefont {Moore}(1999)}]{Koch99}%
  \BibitemOpen
  \bibfield  {author} {\bibinfo {author} {\bibfnamefont {K.}~\bibnamefont
  {Koch}}\ and\ \bibinfo {author} {\bibfnamefont {G.~T.}\ \bibnamefont
  {Moore}},\ }\href@noop {} {\bibfield  {journal} {\bibinfo  {journal} {J.~Opt.
  Soc. Am.~B}\ }\textbf {\bibinfo {volume} {16}},\ \bibinfo {pages} {448}
  (\bibinfo {year} {1999})}\BibitemShut {NoStop}%
\bibitem [{\citenamefont {Liscidini}\ and\ \citenamefont
  {Andreani}(2004)}]{Liscidini04}%
  \BibitemOpen
  \bibfield  {author} {\bibinfo {author} {\bibfnamefont {M.}~\bibnamefont
  {Liscidini}}\ and\ \bibinfo {author} {\bibfnamefont {L.~A.}\ \bibnamefont
  {Andreani}},\ }\href@noop {} {\bibfield  {journal} {\bibinfo  {journal}
  {Appl. Phys. Lett.}\ }\textbf {\bibinfo {volume} {85}},\ \bibinfo {pages}
  {1883} (\bibinfo {year} {2004})}\BibitemShut {NoStop}%
\bibitem [{\citenamefont {Rivoire}\ \emph
  {et~al.}(2011{\natexlab{a}})\citenamefont {Rivoire}, \citenamefont
  {Buckley},\ and\ \citenamefont {Vuckovic}}]{rivoire11:apl}%
  \BibitemOpen
  \bibfield  {author} {\bibinfo {author} {\bibfnamefont {K.}~\bibnamefont
  {Rivoire}}, \bibinfo {author} {\bibfnamefont {S.}~\bibnamefont {Buckley}}, \
  and\ \bibinfo {author} {\bibfnamefont {J.}~\bibnamefont {Vuckovic}},\
  }\href@noop {} {\bibfield  {journal} {\bibinfo  {journal} {Appl. Phys.
  Lett.}\ }\textbf {\bibinfo {volume} {99}},\ \bibinfo {pages} {013114}
  (\bibinfo {year} {2011}{\natexlab{a}})}\BibitemShut {NoStop}%
\bibitem [{\citenamefont {Ramirez}\ \emph {et~al.}(2011)\citenamefont
  {Ramirez}, \citenamefont {Rodriguez}, \citenamefont {Hashemi}, \citenamefont
  {Joannopoulos}, \citenamefont {Solijacic},\ and\ \citenamefont
  {Johnson}}]{Ramirez11}%
  \BibitemOpen
  \bibfield  {author} {\bibinfo {author} {\bibfnamefont {D.}~\bibnamefont
  {Ramirez}}, \bibinfo {author} {\bibfnamefont {A.~W.}\ \bibnamefont
  {Rodriguez}}, \bibinfo {author} {\bibfnamefont {H.}~\bibnamefont {Hashemi}},
  \bibinfo {author} {\bibfnamefont {J.~D.}\ \bibnamefont {Joannopoulos}},
  \bibinfo {author} {\bibfnamefont {M.}~\bibnamefont {Solijacic}}, \ and\
  \bibinfo {author} {\bibfnamefont {S.~G.}\ \bibnamefont {Johnson}},\
  }\href@noop {} {\bibfield  {journal} {\bibinfo  {journal} {Phys. Rev.~A}\
  }\textbf {\bibinfo {volume} {83}},\ \bibinfo {pages} {033834} (\bibinfo
  {year} {2011})}\BibitemShut {NoStop}%
\bibitem [{\citenamefont {Lin}\ \emph {et~al.}(2014)\citenamefont {Lin},
  \citenamefont {Alcorn}, \citenamefont {Loncar}, \citenamefont {Johnson},\
  and\ \citenamefont {Rodriguez}}]{Zin14}%
  \BibitemOpen
  \bibfield  {author} {\bibinfo {author} {\bibfnamefont {Z.}~\bibnamefont
  {Lin}}, \bibinfo {author} {\bibfnamefont {T.}~\bibnamefont {Alcorn}},
  \bibinfo {author} {\bibfnamefont {M.}~\bibnamefont {Loncar}}, \bibinfo
  {author} {\bibfnamefont {S.}~\bibnamefont {Johnson}}, \ and\ \bibinfo
  {author} {\bibfnamefont {A.}~\bibnamefont {Rodriguez}},\ }\href@noop {}
  {\bibfield  {journal} {\bibinfo  {journal} {Phys. Rev.~A}\ }\textbf {\bibinfo
  {volume} {89}},\ \bibinfo {pages} {053839} (\bibinfo {year}
  {2014})}\BibitemShut {NoStop}%
\bibitem [{\citenamefont {Pernice}\ \emph {et~al.}(2012)\citenamefont
  {Pernice}, \citenamefont {Xiong}, \citenamefont {Schuck},\ and\ \citenamefont
  {Tang}}]{Pernice12}%
  \BibitemOpen
  \bibfield  {author} {\bibinfo {author} {\bibfnamefont {W.~H.~P.}\
  \bibnamefont {Pernice}}, \bibinfo {author} {\bibfnamefont {C.}~\bibnamefont
  {Xiong}}, \bibinfo {author} {\bibfnamefont {C.}~\bibnamefont {Schuck}}, \
  and\ \bibinfo {author} {\bibfnamefont {H.~X.}\ \bibnamefont {Tang}},\ }\href
  {\doibase http://dx.doi.org/10.1063/1.4722941} {\bibfield  {journal}
  {\bibinfo  {journal} {Applied Physics Letters}\ }\textbf {\bibinfo {volume}
  {100}},\ \bibinfo {eid} {223501} (\bibinfo {year} {2012}),\
  http://dx.doi.org/10.1063/1.4722941}\BibitemShut {NoStop}%
\bibitem [{\citenamefont {Bi}\ \emph {et~al.}(2012)\citenamefont {Bi},
  \citenamefont {Rodriguez}, \citenamefont {Hashemi}, \citenamefont {Duchesne},
  \citenamefont {Loncar}, \citenamefont {Wang},\ and\ \citenamefont
  {Johnson}}]{Bi12}%
  \BibitemOpen
  \bibfield  {author} {\bibinfo {author} {\bibfnamefont {Z.-F.}\ \bibnamefont
  {Bi}}, \bibinfo {author} {\bibfnamefont {A.~W.}\ \bibnamefont {Rodriguez}},
  \bibinfo {author} {\bibfnamefont {H.}~\bibnamefont {Hashemi}}, \bibinfo
  {author} {\bibfnamefont {D.}~\bibnamefont {Duchesne}}, \bibinfo {author}
  {\bibfnamefont {M.}~\bibnamefont {Loncar}}, \bibinfo {author} {\bibfnamefont
  {K.-M.}\ \bibnamefont {Wang}}, \ and\ \bibinfo {author} {\bibfnamefont
  {S.~G.}\ \bibnamefont {Johnson}},\ }\href {\doibase 10.1364/OE.20.007526}
  {\bibfield  {journal} {\bibinfo  {journal} {Opt. Express}\ }\textbf {\bibinfo
  {volume} {20}},\ \bibinfo {pages} {7526} (\bibinfo {year}
  {2012})}\BibitemShut {NoStop}%
\bibitem [{\citenamefont {Rivoire}\ \emph {et~al.}(2009)\citenamefont
  {Rivoire}, \citenamefont {Lin}, \citenamefont {Hatami}, \citenamefont
  {Masselink},\ and\ \citenamefont {Vu\v{c}kovi\'{c}}}]{Rivoire09}%
  \BibitemOpen
  \bibfield  {author} {\bibinfo {author} {\bibfnamefont {K.}~\bibnamefont
  {Rivoire}}, \bibinfo {author} {\bibfnamefont {Z.}~\bibnamefont {Lin}},
  \bibinfo {author} {\bibfnamefont {F.}~\bibnamefont {Hatami}}, \bibinfo
  {author} {\bibfnamefont {W.~T.}\ \bibnamefont {Masselink}}, \ and\ \bibinfo
  {author} {\bibfnamefont {J.}~\bibnamefont {Vu\v{c}kovi\'{c}}},\ }\href
  {\doibase 10.1364/OE.17.022609} {\bibfield  {journal} {\bibinfo  {journal}
  {Opt. Express}\ }\textbf {\bibinfo {volume} {17}},\ \bibinfo {pages} {22609}
  (\bibinfo {year} {2009})}\BibitemShut {NoStop}%
\bibitem [{\citenamefont {Buckley}\ \emph {et~al.}(2014)\citenamefont
  {Buckley}, \citenamefont {Radulaski}, \citenamefont {Zhang}, \citenamefont
  {Petykiewicz}, \citenamefont {Biermann},\ and\ \citenamefont
  {Vu\v{c}kovi\'{c}}}]{Buckley14}%
  \BibitemOpen
  \bibfield  {author} {\bibinfo {author} {\bibfnamefont {S.}~\bibnamefont
  {Buckley}}, \bibinfo {author} {\bibfnamefont {M.}~\bibnamefont {Radulaski}},
  \bibinfo {author} {\bibfnamefont {J.~L.}\ \bibnamefont {Zhang}}, \bibinfo
  {author} {\bibfnamefont {J.}~\bibnamefont {Petykiewicz}}, \bibinfo {author}
  {\bibfnamefont {K.}~\bibnamefont {Biermann}}, \ and\ \bibinfo {author}
  {\bibfnamefont {J.}~\bibnamefont {Vu\v{c}kovi\'{c}}},\ }\href {\doibase
  10.1364/OE.22.026498} {\bibfield  {journal} {\bibinfo  {journal} {Opt.
  Express}\ }\textbf {\bibinfo {volume} {22}},\ \bibinfo {pages} {26498}
  (\bibinfo {year} {2014})}\BibitemShut {NoStop}%
\bibitem [{\citenamefont {Rodriguez}\ \emph {et~al.}(2007)\citenamefont
  {Rodriguez}, \citenamefont {Solja{\v{c}}i{\'{c}}}, \citenamefont
  {Joannopulos},\ and\ \citenamefont {Johnson}}]{Rodriguez07:OE}%
  \BibitemOpen
  \bibfield  {author} {\bibinfo {author} {\bibfnamefont {A.}~\bibnamefont
  {Rodriguez}}, \bibinfo {author} {\bibfnamefont {M.}~\bibnamefont
  {Solja{\v{c}}i{\'{c}}}}, \bibinfo {author} {\bibfnamefont {J.~D.}\
  \bibnamefont {Joannopulos}}, \ and\ \bibinfo {author} {\bibfnamefont {S.~G.}\
  \bibnamefont {Johnson}},\ }\href@noop {} {\bibfield  {journal} {\bibinfo
  {journal} {Opt. Express}\ }\textbf {\bibinfo {volume} {15}},\ \bibinfo
  {pages} {7303} (\bibinfo {year} {2007})}\BibitemShut {NoStop}%
\bibitem [{\citenamefont {Lin}\ \emph {et~al.}(2016)\citenamefont {Lin},
  \citenamefont {Liang}, \citenamefont {Lon\v{c}ar}, \citenamefont {Johnson},\
  and\ \citenamefont {Rodriguez}}]{Lin:16}%
  \BibitemOpen
  \bibfield  {author} {\bibinfo {author} {\bibfnamefont {Z.}~\bibnamefont
  {Lin}}, \bibinfo {author} {\bibfnamefont {X.}~\bibnamefont {Liang}}, \bibinfo
  {author} {\bibfnamefont {M.}~\bibnamefont {Lon\v{c}ar}}, \bibinfo {author}
  {\bibfnamefont {S.~G.}\ \bibnamefont {Johnson}}, \ and\ \bibinfo {author}
  {\bibfnamefont {A.~W.}\ \bibnamefont {Rodriguez}},\ }\href {\doibase
  10.1364/OPTICA.3.000233} {\bibfield  {journal} {\bibinfo  {journal} {Optica}\
  }\textbf {\bibinfo {volume} {3}},\ \bibinfo {pages} {233} (\bibinfo {year}
  {2016})}\BibitemShut {NoStop}%
\bibitem [{\citenamefont {Lin}\ \emph {et~al.}(2017)\citenamefont {Lin},
  \citenamefont {Lon{\v{c}}ar},\ and\ \citenamefont {Rodriguez}}]{ZinOL}%
  \BibitemOpen
  \bibfield  {author} {\bibinfo {author} {\bibfnamefont {Z.}~\bibnamefont
  {Lin}}, \bibinfo {author} {\bibfnamefont {M.}~\bibnamefont {Lon{\v{c}}ar}}, \
  and\ \bibinfo {author} {\bibfnamefont {A.~W.}\ \bibnamefont {Rodriguez}},\
  }\href@noop {} {\bibfield  {journal} {\bibinfo  {journal} {arXiv preprint
  arXiv:1701.05628}\ } (\bibinfo {year} {2017})}\BibitemShut {NoStop}%
\bibitem [{\citenamefont {Goldberg}\ and\ \citenamefont
  {Kliner}(1995)}]{Lew95}%
  \BibitemOpen
  \bibfield  {author} {\bibinfo {author} {\bibfnamefont {L.}~\bibnamefont
  {Goldberg}}\ and\ \bibinfo {author} {\bibfnamefont {D.~A.~V.}\ \bibnamefont
  {Kliner}},\ }\href@noop {} {\bibfield  {journal} {\bibinfo  {journal} {Opt.
  Lett.}\ }\textbf {\bibinfo {volume} {20}},\ \bibinfo {pages} {1640} (\bibinfo
  {year} {1995})}\BibitemShut {NoStop}%
\bibitem [{\citenamefont {Yelin}(1999)}]{Yelin99}%
  \BibitemOpen
  \bibfield  {author} {\bibinfo {author} {\bibfnamefont {Y.}~\bibnamefont
  {Yelin}, \bibfnamefont {D.and~Silberberg}},\ }\href@noop {} {\bibfield
  {journal} {\bibinfo  {journal} {Opt. Express}\ }\textbf {\bibinfo {volume}
  {5}},\ \bibinfo {pages} {169} (\bibinfo {year} {1999})}\BibitemShut {NoStop}%
\bibitem [{\citenamefont {Pantazis}\ \emph {et~al.}(2010)\citenamefont
  {Pantazis}, \citenamefont {Maloney}, \citenamefont {Wu},\ and\ \citenamefont
  {Fraser}}]{SHGMicroscpy}%
  \BibitemOpen
  \bibfield  {author} {\bibinfo {author} {\bibfnamefont {P.}~\bibnamefont
  {Pantazis}}, \bibinfo {author} {\bibfnamefont {J.}~\bibnamefont {Maloney}},
  \bibinfo {author} {\bibfnamefont {D.}~\bibnamefont {Wu}}, \ and\ \bibinfo
  {author} {\bibfnamefont {S.~E.}\ \bibnamefont {Fraser}},\ }\href@noop {}
  {\bibfield  {journal} {\bibinfo  {journal} {Proceedings of the National
  Academy of Sciences}\ }\textbf {\bibinfo {volume} {107}},\ \bibinfo {pages}
  {14535} (\bibinfo {year} {2010})}\BibitemShut {NoStop}%
\bibitem [{\citenamefont {Hamel}\ \emph {et~al.}(2010)\citenamefont {Hamel},
  \citenamefont {Fedrizzi}, \citenamefont {Ramelow}, \citenamefont {Resch},\
  and\ \citenamefont {Jennewein}}]{Hamel10}%
  \BibitemOpen
  \bibfield  {author} {\bibinfo {author} {\bibfnamefont {D.~R.}\ \bibnamefont
  {Hamel}}, \bibinfo {author} {\bibfnamefont {A.}~\bibnamefont {Fedrizzi}},
  \bibinfo {author} {\bibfnamefont {S.}~\bibnamefont {Ramelow}}, \bibinfo
  {author} {\bibfnamefont {K.~J.}\ \bibnamefont {Resch}}, \ and\ \bibinfo
  {author} {\bibfnamefont {T.}~\bibnamefont {Jennewein}},\ }\href@noop {}
  {\bibfield  {journal} {\bibinfo  {journal} {Nature}\ }\textbf {\bibinfo
  {volume} {466}},\ \bibinfo {pages} {601} (\bibinfo {year}
  {2010})}\BibitemShut {NoStop}%
\bibitem [{\citenamefont {Rivoire}\ \emph
  {et~al.}(2011{\natexlab{b}})\citenamefont {Rivoire}, \citenamefont {Buckley},
  \citenamefont {Hatami},\ and\ \citenamefont {Vuckovic}}]{rivoire11b:apl}%
  \BibitemOpen
  \bibfield  {author} {\bibinfo {author} {\bibfnamefont {K.}~\bibnamefont
  {Rivoire}}, \bibinfo {author} {\bibfnamefont {S.}~\bibnamefont {Buckley}},
  \bibinfo {author} {\bibfnamefont {F.}~\bibnamefont {Hatami}}, \ and\ \bibinfo
  {author} {\bibfnamefont {J.}~\bibnamefont {Vuckovic}},\ }\href@noop {}
  {\bibfield  {journal} {\bibinfo  {journal} {Appl. Phys. Lett.}\ }\textbf
  {\bibinfo {volume} {98}},\ \bibinfo {pages} {263113} (\bibinfo {year}
  {2011}{\natexlab{b}})}\BibitemShut {NoStop}%
\bibitem [{\citenamefont {Buckley}\ \emph {et~al.}(2013)\citenamefont
  {Buckley}, \citenamefont {Radulaski}, \citenamefont {Biermann},\ and\
  \citenamefont {Vuckovic}}]{Buckley13}%
  \BibitemOpen
  \bibfield  {author} {\bibinfo {author} {\bibfnamefont {S.}~\bibnamefont
  {Buckley}}, \bibinfo {author} {\bibfnamefont {M.}~\bibnamefont {Radulaski}},
  \bibinfo {author} {\bibfnamefont {K.}~\bibnamefont {Biermann}}, \ and\
  \bibinfo {author} {\bibfnamefont {J.}~\bibnamefont {Vuckovic}},\ }\href@noop
  {} {\bibfield  {journal} {\bibinfo  {journal} {ArXiv:1308.6051v1}\ }
  (\bibinfo {year} {2013})}\BibitemShut {NoStop}%
\bibitem [{\citenamefont {Hall}\ \emph {et~al.}(1989)\citenamefont {Hall},
  \citenamefont {Newhouse}, \citenamefont {Borrelli}, \citenamefont
  {Dumbaugh},\ and\ \citenamefont {Weidman}}]{Hall89}%
  \BibitemOpen
  \bibfield  {author} {\bibinfo {author} {\bibfnamefont {D.~W.}\ \bibnamefont
  {Hall}}, \bibinfo {author} {\bibfnamefont {M.~A.}\ \bibnamefont {Newhouse}},
  \bibinfo {author} {\bibfnamefont {N.~F.}\ \bibnamefont {Borrelli}}, \bibinfo
  {author} {\bibfnamefont {W.~H.}\ \bibnamefont {Dumbaugh}}, \ and\ \bibinfo
  {author} {\bibfnamefont {D.~L.}\ \bibnamefont {Weidman}},\ }\href {\doibase
  http://dx.doi.org/10.1063/1.100697} {\bibfield  {journal} {\bibinfo
  {journal} {Applied Physics Letters}\ }\textbf {\bibinfo {volume} {54}},\
  \bibinfo {pages} {1293} (\bibinfo {year} {1989})}\BibitemShut {NoStop}%
\bibitem [{\citenamefont {Ahmad}\ \emph {et~al.}(2004)\citenamefont {Ahmad},
  \citenamefont {Soljacic}, \citenamefont {Ibanescu}, \citenamefont {Engeness},
  \citenamefont {Skorobogatly}, \citenamefont {Johnson}, \citenamefont
  {Weisberg}, \citenamefont {Fink}, \citenamefont {Pressman}, \citenamefont
  {King} \emph {et~al.}}]{ahmad2004high}%
  \BibitemOpen
  \bibfield  {author} {\bibinfo {author} {\bibfnamefont {R.}~\bibnamefont
  {Ahmad}}, \bibinfo {author} {\bibfnamefont {M.}~\bibnamefont {Soljacic}},
  \bibinfo {author} {\bibfnamefont {M.}~\bibnamefont {Ibanescu}}, \bibinfo
  {author} {\bibfnamefont {T.}~\bibnamefont {Engeness}}, \bibinfo {author}
  {\bibfnamefont {M.}~\bibnamefont {Skorobogatly}}, \bibinfo {author}
  {\bibfnamefont {S.}~\bibnamefont {Johnson}}, \bibinfo {author} {\bibfnamefont
  {O.}~\bibnamefont {Weisberg}}, \bibinfo {author} {\bibfnamefont
  {Y.}~\bibnamefont {Fink}}, \bibinfo {author} {\bibfnamefont {L.}~\bibnamefont
  {Pressman}}, \bibinfo {author} {\bibfnamefont {W.}~\bibnamefont {King}},
  \emph {et~al.},\ }\href {https://www.google.com/patents/US6788864} {\enquote
  {\bibinfo {title} {High index-contrast fiber waveguides and applications},}\
  } (\bibinfo {year} {2004}),\ \bibinfo {note} {uS Patent
  6,788,864}\BibitemShut {NoStop}%
\bibitem [{\citenamefont {Lapine}\ \emph {et~al.}(2014)\citenamefont {Lapine},
  \citenamefont {Shadrivov},\ and\ \citenamefont
  {Kivshar}}]{lapine2014colloquium}%
  \BibitemOpen
  \bibfield  {author} {\bibinfo {author} {\bibfnamefont {M.}~\bibnamefont
  {Lapine}}, \bibinfo {author} {\bibfnamefont {I.~V.}\ \bibnamefont
  {Shadrivov}}, \ and\ \bibinfo {author} {\bibfnamefont {Y.~S.}\ \bibnamefont
  {Kivshar}},\ }\href@noop {} {\bibfield  {journal} {\bibinfo  {journal}
  {Reviews of Modern Physics}\ }\textbf {\bibinfo {volume} {86}},\ \bibinfo
  {pages} {1093} (\bibinfo {year} {2014})}\BibitemShut {NoStop}%
\bibitem [{\citenamefont {Campione}\ \emph {et~al.}(2014)\citenamefont
  {Campione}, \citenamefont {Benz}, \citenamefont {Sinclair}, \citenamefont
  {Capolino},\ and\ \citenamefont {Brener}}]{campione2014second}%
  \BibitemOpen
  \bibfield  {author} {\bibinfo {author} {\bibfnamefont {S.}~\bibnamefont
  {Campione}}, \bibinfo {author} {\bibfnamefont {A.}~\bibnamefont {Benz}},
  \bibinfo {author} {\bibfnamefont {M.~B.}\ \bibnamefont {Sinclair}}, \bibinfo
  {author} {\bibfnamefont {F.}~\bibnamefont {Capolino}}, \ and\ \bibinfo
  {author} {\bibfnamefont {I.}~\bibnamefont {Brener}},\ }\href@noop {}
  {\bibfield  {journal} {\bibinfo  {journal} {Applied Physics Letters}\
  }\textbf {\bibinfo {volume} {104}},\ \bibinfo {pages} {131104} (\bibinfo
  {year} {2014})}\BibitemShut {NoStop}%
\bibitem [{\citenamefont {Lee}\ \emph {et~al.}(2014)\citenamefont {Lee},
  \citenamefont {Tymchenko}, \citenamefont {Argyropoulos}, \citenamefont
  {Chen}, \citenamefont {Lu}, \citenamefont {Demmerle}, \citenamefont {Boehm},
  \citenamefont {Amann}, \citenamefont {Alu},\ and\ \citenamefont
  {Belkin}}]{lee2014giant}%
  \BibitemOpen
  \bibfield  {author} {\bibinfo {author} {\bibfnamefont {J.}~\bibnamefont
  {Lee}}, \bibinfo {author} {\bibfnamefont {M.}~\bibnamefont {Tymchenko}},
  \bibinfo {author} {\bibfnamefont {C.}~\bibnamefont {Argyropoulos}}, \bibinfo
  {author} {\bibfnamefont {P.-Y.}\ \bibnamefont {Chen}}, \bibinfo {author}
  {\bibfnamefont {F.}~\bibnamefont {Lu}}, \bibinfo {author} {\bibfnamefont
  {F.}~\bibnamefont {Demmerle}}, \bibinfo {author} {\bibfnamefont
  {G.}~\bibnamefont {Boehm}}, \bibinfo {author} {\bibfnamefont {M.-C.}\
  \bibnamefont {Amann}}, \bibinfo {author} {\bibfnamefont {A.}~\bibnamefont
  {Alu}}, \ and\ \bibinfo {author} {\bibfnamefont {M.~A.}\ \bibnamefont
  {Belkin}},\ }\href@noop {} {\bibfield  {journal} {\bibinfo  {journal}
  {Nature}\ }\textbf {\bibinfo {volume} {511}},\ \bibinfo {pages} {65}
  (\bibinfo {year} {2014})}\BibitemShut {NoStop}%
\bibitem [{\citenamefont {Wolf}\ \emph
  {et~al.}(2015{\natexlab{a}})\citenamefont {Wolf}, \citenamefont {Campione},
  \citenamefont {Benz}, \citenamefont {Ravikumar}, \citenamefont {Liu},
  \citenamefont {Luk}, \citenamefont {Kadlec}, \citenamefont {Shaner},
  \citenamefont {Klem}, \citenamefont {Sinclair} \emph
  {et~al.}}]{wolf2015phased}%
  \BibitemOpen
  \bibfield  {author} {\bibinfo {author} {\bibfnamefont {O.}~\bibnamefont
  {Wolf}}, \bibinfo {author} {\bibfnamefont {S.}~\bibnamefont {Campione}},
  \bibinfo {author} {\bibfnamefont {A.}~\bibnamefont {Benz}}, \bibinfo {author}
  {\bibfnamefont {A.~P.}\ \bibnamefont {Ravikumar}}, \bibinfo {author}
  {\bibfnamefont {S.}~\bibnamefont {Liu}}, \bibinfo {author} {\bibfnamefont
  {T.~S.}\ \bibnamefont {Luk}}, \bibinfo {author} {\bibfnamefont {E.~A.}\
  \bibnamefont {Kadlec}}, \bibinfo {author} {\bibfnamefont {E.~A.}\
  \bibnamefont {Shaner}}, \bibinfo {author} {\bibfnamefont {J.~F.}\
  \bibnamefont {Klem}}, \bibinfo {author} {\bibfnamefont {M.~B.}\ \bibnamefont
  {Sinclair}},  \emph {et~al.},\ }\href@noop {} {\bibfield  {journal} {\bibinfo
   {journal} {Nature communications}\ }\textbf {\bibinfo {volume} {6}}
  (\bibinfo {year} {2015}{\natexlab{a}})}\BibitemShut {NoStop}%
\bibitem [{\citenamefont {O~Brien}\ \emph {et~al.}(2015)\citenamefont
  {O~Brien}, \citenamefont {Suchowski}, \citenamefont {Rho}, \citenamefont
  {Salandrino}, \citenamefont {Kante}, \citenamefont {Yin},\ and\ \citenamefont
  {Zhang}}]{o2015predicting}%
  \BibitemOpen
  \bibfield  {author} {\bibinfo {author} {\bibfnamefont {K.}~\bibnamefont
  {O~Brien}}, \bibinfo {author} {\bibfnamefont {H.}~\bibnamefont {Suchowski}},
  \bibinfo {author} {\bibfnamefont {J.}~\bibnamefont {Rho}}, \bibinfo {author}
  {\bibfnamefont {A.}~\bibnamefont {Salandrino}}, \bibinfo {author}
  {\bibfnamefont {B.}~\bibnamefont {Kante}}, \bibinfo {author} {\bibfnamefont
  {X.}~\bibnamefont {Yin}}, \ and\ \bibinfo {author} {\bibfnamefont
  {X.}~\bibnamefont {Zhang}},\ }\href@noop {} {\bibfield  {journal} {\bibinfo
  {journal} {Nature materials}\ }\textbf {\bibinfo {volume} {14}},\ \bibinfo
  {pages} {379} (\bibinfo {year} {2015})}\BibitemShut {NoStop}%
\bibitem [{\citenamefont {Yang}\ \emph {et~al.}(2015)\citenamefont {Yang},
  \citenamefont {Wang}, \citenamefont {Boulesbaa}, \citenamefont {Kravchenko},
  \citenamefont {Briggs}, \citenamefont {Puretzky}, \citenamefont {Geohegan},\
  and\ \citenamefont {Valentine}}]{yang2015nonlinear}%
  \BibitemOpen
  \bibfield  {author} {\bibinfo {author} {\bibfnamefont {Y.}~\bibnamefont
  {Yang}}, \bibinfo {author} {\bibfnamefont {W.}~\bibnamefont {Wang}}, \bibinfo
  {author} {\bibfnamefont {A.}~\bibnamefont {Boulesbaa}}, \bibinfo {author}
  {\bibfnamefont {I.~I.}\ \bibnamefont {Kravchenko}}, \bibinfo {author}
  {\bibfnamefont {D.~P.}\ \bibnamefont {Briggs}}, \bibinfo {author}
  {\bibfnamefont {A.}~\bibnamefont {Puretzky}}, \bibinfo {author}
  {\bibfnamefont {D.}~\bibnamefont {Geohegan}}, \ and\ \bibinfo {author}
  {\bibfnamefont {J.}~\bibnamefont {Valentine}},\ }\href@noop {} {\bibfield
  {journal} {\bibinfo  {journal} {Nano letters}\ }\textbf {\bibinfo {volume}
  {15}},\ \bibinfo {pages} {7388} (\bibinfo {year} {2015})}\BibitemShut
  {NoStop}%
\bibitem [{\citenamefont {Segal}\ \emph {et~al.}(2015)\citenamefont {Segal},
  \citenamefont {Keren-Zur}, \citenamefont {Hendler},\ and\ \citenamefont
  {Ellenbogen}}]{segal2015controlling}%
  \BibitemOpen
  \bibfield  {author} {\bibinfo {author} {\bibfnamefont {N.}~\bibnamefont
  {Segal}}, \bibinfo {author} {\bibfnamefont {S.}~\bibnamefont {Keren-Zur}},
  \bibinfo {author} {\bibfnamefont {N.}~\bibnamefont {Hendler}}, \ and\
  \bibinfo {author} {\bibfnamefont {T.}~\bibnamefont {Ellenbogen}},\
  }\href@noop {} {\bibfield  {journal} {\bibinfo  {journal} {Nature Photonics}\
  }\textbf {\bibinfo {volume} {9}},\ \bibinfo {pages} {180} (\bibinfo {year}
  {2015})}\BibitemShut {NoStop}%
\bibitem [{\citenamefont {Butet}\ \emph {et~al.}(2015)\citenamefont {Butet},
  \citenamefont {Brevet},\ and\ \citenamefont {Martin}}]{butet2015optical}%
  \BibitemOpen
  \bibfield  {author} {\bibinfo {author} {\bibfnamefont {J.}~\bibnamefont
  {Butet}}, \bibinfo {author} {\bibfnamefont {P.-F.}\ \bibnamefont {Brevet}}, \
  and\ \bibinfo {author} {\bibfnamefont {O.~J.}\ \bibnamefont {Martin}},\
  }\href@noop {} {\bibfield  {journal} {\bibinfo  {journal} {ACS nano}\
  }\textbf {\bibinfo {volume} {9}},\ \bibinfo {pages} {10545} (\bibinfo {year}
  {2015})}\BibitemShut {NoStop}%
\bibitem [{\citenamefont {B\'{e}tourn\'{e}}\ \emph {et~al.}(2008)\citenamefont
  {B\'{e}tourn\'{e}}, \citenamefont {Quiquempois}, \citenamefont {Bouwmans},\
  and\ \citenamefont {Douay}}]{Betourne08}%
  \BibitemOpen
  \bibfield  {author} {\bibinfo {author} {\bibfnamefont {A.}~\bibnamefont
  {B\'{e}tourn\'{e}}}, \bibinfo {author} {\bibfnamefont {Y.}~\bibnamefont
  {Quiquempois}}, \bibinfo {author} {\bibfnamefont {G.}~\bibnamefont
  {Bouwmans}}, \ and\ \bibinfo {author} {\bibfnamefont {M.}~\bibnamefont
  {Douay}},\ }\href {\doibase 10.1364/OE.16.014255} {\bibfield  {journal}
  {\bibinfo  {journal} {Opt. Express}\ }\textbf {\bibinfo {volume} {16}},\
  \bibinfo {pages} {14255} (\bibinfo {year} {2008})}\BibitemShut {NoStop}%
\bibitem [{\citenamefont {Li}\ \emph {et~al.}(2017)\citenamefont {Li},
  \citenamefont {Zhang},\ and\ \citenamefont {Zentgraf}}]{li2017nonlinear}%
  \BibitemOpen
  \bibfield  {author} {\bibinfo {author} {\bibfnamefont {G.}~\bibnamefont
  {Li}}, \bibinfo {author} {\bibfnamefont {S.}~\bibnamefont {Zhang}}, \ and\
  \bibinfo {author} {\bibfnamefont {T.}~\bibnamefont {Zentgraf}},\ }\href@noop
  {} {\bibfield  {journal} {\bibinfo  {journal} {Nature Reviews Materials}\
  }\textbf {\bibinfo {volume} {2}},\ \bibinfo {pages} {natrevmats201710}
  (\bibinfo {year} {2017})}\BibitemShut {NoStop}%
\bibitem [{\citenamefont {Liu}\ \emph {et~al.}(2016)\citenamefont {Liu},
  \citenamefont {Sinclair}, \citenamefont {Saravi}, \citenamefont {Keeler},
  \citenamefont {Yang}, \citenamefont {Reno}, \citenamefont {Peake},
  \citenamefont {Setzpfandt}, \citenamefont {Staude}, \citenamefont {Pertsch}
  \emph {et~al.}}]{liu2016resonantly}%
  \BibitemOpen
  \bibfield  {author} {\bibinfo {author} {\bibfnamefont {S.}~\bibnamefont
  {Liu}}, \bibinfo {author} {\bibfnamefont {M.~B.}\ \bibnamefont {Sinclair}},
  \bibinfo {author} {\bibfnamefont {S.}~\bibnamefont {Saravi}}, \bibinfo
  {author} {\bibfnamefont {G.~A.}\ \bibnamefont {Keeler}}, \bibinfo {author}
  {\bibfnamefont {Y.}~\bibnamefont {Yang}}, \bibinfo {author} {\bibfnamefont
  {J.}~\bibnamefont {Reno}}, \bibinfo {author} {\bibfnamefont {G.~M.}\
  \bibnamefont {Peake}}, \bibinfo {author} {\bibfnamefont {F.}~\bibnamefont
  {Setzpfandt}}, \bibinfo {author} {\bibfnamefont {I.}~\bibnamefont {Staude}},
  \bibinfo {author} {\bibfnamefont {T.}~\bibnamefont {Pertsch}},  \emph
  {et~al.},\ }\href@noop {} {\bibfield  {journal} {\bibinfo  {journal} {Nano
  letters}\ }\textbf {\bibinfo {volume} {16}},\ \bibinfo {pages} {5426}
  (\bibinfo {year} {2016})}\BibitemShut {NoStop}%
\bibitem [{\citenamefont {Wolf}\ \emph
  {et~al.}(2015{\natexlab{b}})\citenamefont {Wolf}, \citenamefont {Allerman},
  \citenamefont {Ma}, \citenamefont {Wendt}, \citenamefont {Song},
  \citenamefont {Shaner},\ and\ \citenamefont {Brener}}]{wolf2015enhanced}%
  \BibitemOpen
  \bibfield  {author} {\bibinfo {author} {\bibfnamefont {O.}~\bibnamefont
  {Wolf}}, \bibinfo {author} {\bibfnamefont {A.~A.}\ \bibnamefont {Allerman}},
  \bibinfo {author} {\bibfnamefont {X.}~\bibnamefont {Ma}}, \bibinfo {author}
  {\bibfnamefont {J.~R.}\ \bibnamefont {Wendt}}, \bibinfo {author}
  {\bibfnamefont {A.~Y.}\ \bibnamefont {Song}}, \bibinfo {author}
  {\bibfnamefont {E.~A.}\ \bibnamefont {Shaner}}, \ and\ \bibinfo {author}
  {\bibfnamefont {I.}~\bibnamefont {Brener}},\ }\href@noop {} {\bibfield
  {journal} {\bibinfo  {journal} {Applied Physics Letters}\ }\textbf {\bibinfo
  {volume} {107}},\ \bibinfo {pages} {151108} (\bibinfo {year}
  {2015}{\natexlab{b}})}\BibitemShut {NoStop}%
\bibitem [{\citenamefont {Tymchenko}\ \emph {et~al.}(2015)\citenamefont
  {Tymchenko}, \citenamefont {Gomez-Diaz}, \citenamefont {Lee}, \citenamefont
  {Nookala}, \citenamefont {Belkin},\ and\ \citenamefont
  {Al{\`u}}}]{tymchenko2015gradient}%
  \BibitemOpen
  \bibfield  {author} {\bibinfo {author} {\bibfnamefont {M.}~\bibnamefont
  {Tymchenko}}, \bibinfo {author} {\bibfnamefont {J.~S.}\ \bibnamefont
  {Gomez-Diaz}}, \bibinfo {author} {\bibfnamefont {J.}~\bibnamefont {Lee}},
  \bibinfo {author} {\bibfnamefont {N.}~\bibnamefont {Nookala}}, \bibinfo
  {author} {\bibfnamefont {M.~A.}\ \bibnamefont {Belkin}}, \ and\ \bibinfo
  {author} {\bibfnamefont {A.}~\bibnamefont {Al{\`u}}},\ }\href@noop {}
  {\bibfield  {journal} {\bibinfo  {journal} {Physical review letters}\
  }\textbf {\bibinfo {volume} {115}},\ \bibinfo {pages} {207403} (\bibinfo
  {year} {2015})}\BibitemShut {NoStop}%
\bibitem [{\citenamefont {Joannopoulos}\ \emph {et~al.}(2008)\citenamefont
  {Joannopoulos}, \citenamefont {Johnson}, \citenamefont {Winn},\ and\
  \citenamefont {Meade}}]{JoannopoulosJo08-book}%
  \BibitemOpen
  \bibfield  {author} {\bibinfo {author} {\bibfnamefont {J.~D.}\ \bibnamefont
  {Joannopoulos}}, \bibinfo {author} {\bibfnamefont {S.~G.}\ \bibnamefont
  {Johnson}}, \bibinfo {author} {\bibfnamefont {J.~N.}\ \bibnamefont {Winn}}, \
  and\ \bibinfo {author} {\bibfnamefont {R.~D.}\ \bibnamefont {Meade}},\ }\href
  {http://ab-initio.mit.edu/book} {\emph {\bibinfo {title} {Photonic Crystals:
  Molding the Flow of Light}}},\ \bibinfo {edition} {2nd}\ ed.\ (\bibinfo
  {publisher} {Princeton University Press},\ \bibinfo {year}
  {2008})\BibitemShut {NoStop}%
\bibitem [{\citenamefont {Kim}\ and\ \citenamefont {O'Brien}(2004)}]{Kim:04}%
  \BibitemOpen
  \bibfield  {author} {\bibinfo {author} {\bibfnamefont {W.~J.}\ \bibnamefont
  {Kim}}\ and\ \bibinfo {author} {\bibfnamefont {J.~D.}\ \bibnamefont
  {O'Brien}},\ }\href {\doibase 10.1364/JOSAB.21.000289} {\bibfield  {journal}
  {\bibinfo  {journal} {J. Opt. Soc. Am. B}\ }\textbf {\bibinfo {volume}
  {21}},\ \bibinfo {pages} {289} (\bibinfo {year} {2004})}\BibitemShut
  {NoStop}%
\bibitem [{\citenamefont {Darki}\ and\ \citenamefont
  {Granpayeh}(2010)}]{Darki10}%
  \BibitemOpen
  \bibfield  {author} {\bibinfo {author} {\bibfnamefont {B.~S.}\ \bibnamefont
  {Darki}}\ and\ \bibinfo {author} {\bibfnamefont {N.}~\bibnamefont
  {Granpayeh}},\ }\href {\doibase
  http://dx.doi.org/10.1016/j.optcom.2010.06.013} {\bibfield  {journal}
  {\bibinfo  {journal} {Optics Communications}\ }\textbf {\bibinfo {volume}
  {283}},\ \bibinfo {pages} {4099 } (\bibinfo {year} {2010})}\BibitemShut
  {NoStop}%
\bibitem [{\citenamefont {Minkov}\ and\ \citenamefont
  {Savona}(2014)}]{Minkov14}%
  \BibitemOpen
  \bibfield  {author} {\bibinfo {author} {\bibfnamefont {M.}~\bibnamefont
  {Minkov}}\ and\ \bibinfo {author} {\bibfnamefont {V.}~\bibnamefont
  {Savona}},\ }\href {http://dx.doi.org/10.1038/srep05124} {\bibfield
  {journal} {\bibinfo  {journal} {Sci. Rep.}\ }\textbf {\bibinfo {volume} {4}}
  (\bibinfo {year} {2014})}\BibitemShut {NoStop}%
\bibitem [{\citenamefont {Gondarenko}\ \emph {et~al.}(2006)\citenamefont
  {Gondarenko}, \citenamefont {Preble}, \citenamefont {Robinson}, \citenamefont
  {Chen}, \citenamefont {Lipson},\ and\ \citenamefont {Lipson}}]{Chen06}%
  \BibitemOpen
  \bibfield  {author} {\bibinfo {author} {\bibfnamefont {A.}~\bibnamefont
  {Gondarenko}}, \bibinfo {author} {\bibfnamefont {S.}~\bibnamefont {Preble}},
  \bibinfo {author} {\bibfnamefont {J.}~\bibnamefont {Robinson}}, \bibinfo
  {author} {\bibfnamefont {L.}~\bibnamefont {Chen}}, \bibinfo {author}
  {\bibfnamefont {H.}~\bibnamefont {Lipson}}, \ and\ \bibinfo {author}
  {\bibfnamefont {M.}~\bibnamefont {Lipson}},\ }\href {\doibase
  10.1103/PhysRevLett.96.143904} {\bibfield  {journal} {\bibinfo  {journal}
  {Phys. Rev. Lett.}\ }\textbf {\bibinfo {volume} {96}},\ \bibinfo {pages}
  {143904} (\bibinfo {year} {2006})}\BibitemShut {NoStop}%
\bibitem [{\citenamefont {Jensen}\ and\ \citenamefont
  {Sigmund}(2011)}]{Jensen11}%
  \BibitemOpen
  \bibfield  {author} {\bibinfo {author} {\bibfnamefont {J.}~\bibnamefont
  {Jensen}}\ and\ \bibinfo {author} {\bibfnamefont {O.}~\bibnamefont
  {Sigmund}},\ }\href {\doibase 10.1002/lpor.201000014} {\bibfield  {journal}
  {\bibinfo  {journal} {Laser and Photonics Reviews}\ }\textbf {\bibinfo
  {volume} {5}},\ \bibinfo {pages} {308} (\bibinfo {year} {2011})}\BibitemShut
  {NoStop}%
\bibitem [{\citenamefont {Liang}\ and\ \citenamefont
  {Johnson}(2013)}]{Liang13}%
  \BibitemOpen
  \bibfield  {author} {\bibinfo {author} {\bibfnamefont {X.}~\bibnamefont
  {Liang}}\ and\ \bibinfo {author} {\bibfnamefont {S.~G.}\ \bibnamefont
  {Johnson}},\ }\href {\doibase 10.1364/OE.21.030812} {\bibfield  {journal}
  {\bibinfo  {journal} {Opt. Express}\ }\textbf {\bibinfo {volume} {21}},\
  \bibinfo {pages} {30812} (\bibinfo {year} {2013})}\BibitemShut {NoStop}%
\bibitem [{\citenamefont {Liu}\ \emph {et~al.}(2013)\citenamefont {Liu},
  \citenamefont {Gabrielli}, \citenamefont {Lipson},\ and\ \citenamefont
  {Johnson}}]{Liu13}%
  \BibitemOpen
  \bibfield  {author} {\bibinfo {author} {\bibfnamefont {D.}~\bibnamefont
  {Liu}}, \bibinfo {author} {\bibfnamefont {L.~H.}\ \bibnamefont {Gabrielli}},
  \bibinfo {author} {\bibfnamefont {M.}~\bibnamefont {Lipson}}, \ and\ \bibinfo
  {author} {\bibfnamefont {S.~G.}\ \bibnamefont {Johnson}},\ }\href {\doibase
  10.1364/OE.21.014223} {\bibfield  {journal} {\bibinfo  {journal} {Opt.
  Express}\ }\textbf {\bibinfo {volume} {21}},\ \bibinfo {pages} {14223}
  (\bibinfo {year} {2013})}\BibitemShut {NoStop}%
\bibitem [{\citenamefont {Piggott}\ \emph {et~al.}(2014)\citenamefont
  {Piggott}, \citenamefont {Lu}, \citenamefont {Babinec}, \citenamefont
  {Lagoudakis}, \citenamefont {Petykiewicz},\ and\ \citenamefont
  {Vuckovic}}]{Piggott14}%
  \BibitemOpen
  \bibfield  {author} {\bibinfo {author} {\bibfnamefont {A.~Y.}\ \bibnamefont
  {Piggott}}, \bibinfo {author} {\bibfnamefont {J.}~\bibnamefont {Lu}},
  \bibinfo {author} {\bibfnamefont {T.~M.}\ \bibnamefont {Babinec}}, \bibinfo
  {author} {\bibfnamefont {K.~G.}\ \bibnamefont {Lagoudakis}}, \bibinfo
  {author} {\bibfnamefont {J.}~\bibnamefont {Petykiewicz}}, \ and\ \bibinfo
  {author} {\bibfnamefont {J.}~\bibnamefont {Vuckovic}},\ }\href
  {http://dx.doi.org/10.1038/srep07210} {\bibfield  {journal} {\bibinfo
  {journal} {Sci. Rep.}\ }\textbf {\bibinfo {volume} {4}} (\bibinfo {year}
  {2014})}\BibitemShut {NoStop}%
\bibitem [{\citenamefont {Men}\ \emph {et~al.}(2014)\citenamefont {Men},
  \citenamefont {Lee}, \citenamefont {Freund}, \citenamefont {Peraire},\ and\
  \citenamefont {Johnson}}]{MenLee14}%
  \BibitemOpen
  \bibfield  {author} {\bibinfo {author} {\bibfnamefont {H.}~\bibnamefont
  {Men}}, \bibinfo {author} {\bibfnamefont {K.~Y.~K.}\ \bibnamefont {Lee}},
  \bibinfo {author} {\bibfnamefont {R.~M.}\ \bibnamefont {Freund}}, \bibinfo
  {author} {\bibfnamefont {J.}~\bibnamefont {Peraire}}, \ and\ \bibinfo
  {author} {\bibfnamefont {S.~G.}\ \bibnamefont {Johnson}},\ }\href {\doibase
  doi:10.1364/OE.22.022632} {\bibfield  {journal} {\bibinfo  {journal} {Optics
  Express}\ }\textbf {\bibinfo {volume} {22}},\ \bibinfo {pages} {22632}
  (\bibinfo {year} {2014})},\ \Eprint {http://arxiv.org/abs/arXiv:1405.4350}
  {arXiv:1405.4350} \BibitemShut {NoStop}%
\bibitem [{\citenamefont {Piggott}\ \emph {et~al.}(2015)\citenamefont
  {Piggott}, \citenamefont {Lu}, \citenamefont {Lagoudakis}, \citenamefont
  {Petykiewicz}, \citenamefont {Babinec},\ and\ \citenamefont
  {Vuckovic}}]{Piggott15}%
  \BibitemOpen
  \bibfield  {author} {\bibinfo {author} {\bibfnamefont {A.~Y.}\ \bibnamefont
  {Piggott}}, \bibinfo {author} {\bibfnamefont {J.}~\bibnamefont {Lu}},
  \bibinfo {author} {\bibfnamefont {K.~G.}\ \bibnamefont {Lagoudakis}},
  \bibinfo {author} {\bibfnamefont {J.}~\bibnamefont {Petykiewicz}}, \bibinfo
  {author} {\bibfnamefont {T.~M.}\ \bibnamefont {Babinec}}, \ and\ \bibinfo
  {author} {\bibfnamefont {J.}~\bibnamefont {Vuckovic}},\ }\href@noop {}
  {\bibfield  {journal} {\bibinfo  {journal} {Nature Photonics}\ }\textbf
  {\bibinfo {volume} {9}},\ \bibinfo {pages} {374 } (\bibinfo {year}
  {2015})}\BibitemShut {NoStop}%
\bibitem [{\citenamefont {Shen}\ \emph {et~al.}(2015)\citenamefont {Shen},
  \citenamefont {Wang},\ and\ \citenamefont {Menon}}]{Shen15}%
  \BibitemOpen
  \bibfield  {author} {\bibinfo {author} {\bibfnamefont {B.}~\bibnamefont
  {Shen}}, \bibinfo {author} {\bibfnamefont {P.}~\bibnamefont {Wang}}, \ and\
  \bibinfo {author} {\bibfnamefont {R.}~\bibnamefont {Menon}},\ }\href@noop {}
  {\bibfield  {journal} {\bibinfo  {journal} {Nature Photonics}\ }\textbf
  {\bibinfo {volume} {9}},\ \bibinfo {pages} {378 } (\bibinfo {year}
  {2015})}\BibitemShut {NoStop}%
\bibitem [{\citenamefont {Strang}(2007)}]{StrangComp}%
  \BibitemOpen
  \bibfield  {author} {\bibinfo {author} {\bibfnamefont {G.}~\bibnamefont
  {Strang}},\ }\href@noop {} {\emph {\bibinfo {title} {Computational science
  and engineering}}},\ Vol.\ \bibinfo {volume} {791}\ (\bibinfo  {publisher}
  {Wellesley-Cambridge Press Wellesley},\ \bibinfo {year} {2007})\BibitemShut
  {NoStop}%
\bibitem [{\citenamefont {Deaton}\ and\ \citenamefont
  {Grandhi}(2014)}]{deaton2014survey}%
  \BibitemOpen
  \bibfield  {author} {\bibinfo {author} {\bibfnamefont {J.~D.}\ \bibnamefont
  {Deaton}}\ and\ \bibinfo {author} {\bibfnamefont {R.~V.}\ \bibnamefont
  {Grandhi}},\ }\href@noop {} {\bibfield  {journal} {\bibinfo  {journal}
  {Structural and Multidisciplinary Optimization}\ }\textbf {\bibinfo {volume}
  {49}},\ \bibinfo {pages} {1} (\bibinfo {year} {2014})}\BibitemShut {NoStop}%
\bibitem [{\citenamefont {Bends{\o}e}\ \emph {et~al.}(2004)\citenamefont
  {Bends{\o}e}, \citenamefont {Sigmund}, \citenamefont {Bends{\o}e},\ and\
  \citenamefont {Sigmund}}]{bendsoe2004topology}%
  \BibitemOpen
  \bibfield  {author} {\bibinfo {author} {\bibfnamefont {M.~P.}\ \bibnamefont
  {Bends{\o}e}}, \bibinfo {author} {\bibfnamefont {O.}~\bibnamefont {Sigmund}},
  \bibinfo {author} {\bibfnamefont {M.~P.}\ \bibnamefont {Bends{\o}e}}, \ and\
  \bibinfo {author} {\bibfnamefont {O.}~\bibnamefont {Sigmund}},\ }\href@noop
  {} {\emph {\bibinfo {title} {Topology optimization by distribution of
  isotropic material}}}\ (\bibinfo  {publisher} {Springer},\ \bibinfo {year}
  {2004})\BibitemShut {NoStop}%
\bibitem [{\citenamefont {Wang}\ \emph {et~al.}(2003)\citenamefont {Wang},
  \citenamefont {Wang},\ and\ \citenamefont {Guo}}]{LvS}%
  \BibitemOpen
  \bibfield  {author} {\bibinfo {author} {\bibfnamefont {M.~Y.}\ \bibnamefont
  {Wang}}, \bibinfo {author} {\bibfnamefont {X.}~\bibnamefont {Wang}}, \ and\
  \bibinfo {author} {\bibfnamefont {D.}~\bibnamefont {Guo}},\ }\href@noop {}
  {\bibfield  {journal} {\bibinfo  {journal} {Computer methods in applied
  mechanics and engineering}\ }\textbf {\bibinfo {volume} {192}},\ \bibinfo
  {pages} {227} (\bibinfo {year} {2003})}\BibitemShut {NoStop}%
\bibitem [{\citenamefont {Haslinger}\ and\ \citenamefont
  {M{\"a}kinen}(2003)}]{haslinger2003introduction}%
  \BibitemOpen
  \bibfield  {author} {\bibinfo {author} {\bibfnamefont {J.}~\bibnamefont
  {Haslinger}}\ and\ \bibinfo {author} {\bibfnamefont {R.~A.}\ \bibnamefont
  {M{\"a}kinen}},\ }\href@noop {} {\emph {\bibinfo {title} {Introduction to
  shape optimization: theory, approximation, and computation}}}\ (\bibinfo
  {publisher} {SIAM},\ \bibinfo {year} {2003})\BibitemShut {NoStop}%
\bibitem [{\citenamefont {Svanberg}(2002)}]{Svanberg02}%
  \BibitemOpen
  \bibfield  {author} {\bibinfo {author} {\bibfnamefont {K.}~\bibnamefont
  {Svanberg}},\ }\href@noop {} {\bibfield  {journal} {\bibinfo  {journal} {SIAM
  Journal on Optimization}\ ,\ \bibinfo {pages} {555}} (\bibinfo {year}
  {2002})}\BibitemShut {NoStop}%
\bibitem [{\citenamefont {Wang}\ and\ \citenamefont
  {Sigmund}(2017)}]{wang2017optimization}%
  \BibitemOpen
  \bibfield  {author} {\bibinfo {author} {\bibfnamefont {F.}~\bibnamefont
  {Wang}}\ and\ \bibinfo {author} {\bibfnamefont {O.}~\bibnamefont {Sigmund}},\
  }in\ \href@noop {} {\emph {\bibinfo {booktitle} {Numerical Simulation of
  Optoelectronic Devices (NUSOD), 2017 International Conference on}}}\
  (\bibinfo {organization} {IEEE},\ \bibinfo {year} {2017})\ pp.\ \bibinfo
  {pages} {39--40}\BibitemShut {NoStop}%
\bibitem [{\citenamefont {Boyd}(1992)}]{Boyd92}%
  \BibitemOpen
  \bibfield  {author} {\bibinfo {author} {\bibfnamefont {R.~W.}\ \bibnamefont
  {Boyd}},\ }\href@noop {} {\emph {\bibinfo {title} {Nonlinear Optics}}}\
  (\bibinfo  {publisher} {Academic Press},\ \bibinfo {address} {California},\
  \bibinfo {year} {1992})\BibitemShut {NoStop}%
\bibitem [{\citenamefont {Taflove}\ and\ \citenamefont
  {Hagness}(2000)}]{Taflove00}%
  \BibitemOpen
  \bibfield  {author} {\bibinfo {author} {\bibfnamefont {A.}~\bibnamefont
  {Taflove}}\ and\ \bibinfo {author} {\bibfnamefont {S.~C.}\ \bibnamefont
  {Hagness}},\ }\href@noop {} {\emph {\bibinfo {title} {Computational
  Electrodynamics: The Finite-Difference Time-Domain Method}}}\ (\bibinfo
  {publisher} {Artech},\ \bibinfo {address} {Norwood, MA},\ \bibinfo {year}
  {2000})\BibitemShut {NoStop}%
\bibitem [{\citenamefont {Temelkuran}\ \emph {et~al.}(2002)\citenamefont
  {Temelkuran}, \citenamefont {Hart}, \citenamefont {Benoit}, \citenamefont
  {Joannopoulos},\ and\ \citenamefont {Fink}}]{Fink02}%
  \BibitemOpen
  \bibfield  {author} {\bibinfo {author} {\bibfnamefont {B.}~\bibnamefont
  {Temelkuran}}, \bibinfo {author} {\bibfnamefont {S.~D.}\ \bibnamefont
  {Hart}}, \bibinfo {author} {\bibfnamefont {G.}~\bibnamefont {Benoit}},
  \bibinfo {author} {\bibfnamefont {J.~D.}\ \bibnamefont {Joannopoulos}}, \
  and\ \bibinfo {author} {\bibfnamefont {Y.}~\bibnamefont {Fink}},\ }\href@noop
  {} {\bibfield  {journal} {\bibinfo  {journal} {Nature}\ }\textbf {\bibinfo
  {volume} {420}},\ \bibinfo {pages} {650 } (\bibinfo {year}
  {2002})}\BibitemShut {NoStop}%
\bibitem [{\citenamefont {Feng}\ \emph {et~al.}(2003)\citenamefont {Feng},
  \citenamefont {Monro}, \citenamefont {Petropoulos}, \citenamefont {Finazzi},\
  and\ \citenamefont {Hewak}}]{Feng03}%
  \BibitemOpen
  \bibfield  {author} {\bibinfo {author} {\bibfnamefont {X.}~\bibnamefont
  {Feng}}, \bibinfo {author} {\bibfnamefont {T.}~\bibnamefont {Monro}},
  \bibinfo {author} {\bibfnamefont {P.}~\bibnamefont {Petropoulos}}, \bibinfo
  {author} {\bibfnamefont {V.}~\bibnamefont {Finazzi}}, \ and\ \bibinfo
  {author} {\bibfnamefont {D.}~\bibnamefont {Hewak}},\ }\href {\doibase
  10.1364/OE.11.002225} {\bibfield  {journal} {\bibinfo  {journal} {Opt.
  Express}\ }\textbf {\bibinfo {volume} {11}},\ \bibinfo {pages} {2225}
  (\bibinfo {year} {2003})}\BibitemShut {NoStop}%
\bibitem [{\citenamefont {Grubsky}\ and\ \citenamefont
  {Savchenko}(2005)}]{Grubsky:05}%
  \BibitemOpen
  \bibfield  {author} {\bibinfo {author} {\bibfnamefont {V.}~\bibnamefont
  {Grubsky}}\ and\ \bibinfo {author} {\bibfnamefont {A.}~\bibnamefont
  {Savchenko}},\ }\href {\doibase 10.1364/OPEX.13.006798} {\bibfield  {journal}
  {\bibinfo  {journal} {Opt. Express}\ }\textbf {\bibinfo {volume} {13}},\
  \bibinfo {pages} {6798} (\bibinfo {year} {2005})}\BibitemShut {NoStop}%
\bibitem [{\citenamefont {Agrawal}(2012)}]{fiberoptics}%
  \BibitemOpen
  \bibfield  {author} {\bibinfo {author} {\bibfnamefont {G.~P.}\ \bibnamefont
  {Agrawal}},\ }\href@noop {} {\emph {\bibinfo {title} {Fiber-optic
  communication systems}}},\ Vol.\ \bibinfo {volume} {222}\ (\bibinfo
  {publisher} {John Wiley \& Sons},\ \bibinfo {year} {2012})\BibitemShut
  {NoStop}%
\bibitem [{\citenamefont {Yu}\ and\ \citenamefont
  {Capasso}(2014)}]{yu2014flat}%
  \BibitemOpen
  \bibfield  {author} {\bibinfo {author} {\bibfnamefont {N.}~\bibnamefont
  {Yu}}\ and\ \bibinfo {author} {\bibfnamefont {F.}~\bibnamefont {Capasso}},\
  }\href@noop {} {\bibfield  {journal} {\bibinfo  {journal} {Nature materials}\
  }\textbf {\bibinfo {volume} {13}},\ \bibinfo {pages} {139} (\bibinfo {year}
  {2014})}\BibitemShut {NoStop}%
\bibitem [{\citenamefont {Michaeli}\ \emph {et~al.}(2017)\citenamefont
  {Michaeli}, \citenamefont {Keren-Zur}, \citenamefont {Avayu}, \citenamefont
  {Suchowski},\ and\ \citenamefont {Ellenbogen}}]{michaeli2017nonlinear}%
  \BibitemOpen
  \bibfield  {author} {\bibinfo {author} {\bibfnamefont {L.}~\bibnamefont
  {Michaeli}}, \bibinfo {author} {\bibfnamefont {S.}~\bibnamefont {Keren-Zur}},
  \bibinfo {author} {\bibfnamefont {O.}~\bibnamefont {Avayu}}, \bibinfo
  {author} {\bibfnamefont {H.}~\bibnamefont {Suchowski}}, \ and\ \bibinfo
  {author} {\bibfnamefont {T.}~\bibnamefont {Ellenbogen}},\ }\href@noop {}
  {\bibfield  {journal} {\bibinfo  {journal} {Physical Review Letters}\
  }\textbf {\bibinfo {volume} {118}},\ \bibinfo {pages} {243904} (\bibinfo
  {year} {2017})}\BibitemShut {NoStop}%
\bibitem [{\citenamefont {Keren-Zur}\ \emph {et~al.}(2015)\citenamefont
  {Keren-Zur}, \citenamefont {Avayu}, \citenamefont {Michaeli},\ and\
  \citenamefont {Ellenbogen}}]{keren2015nonlinear}%
  \BibitemOpen
  \bibfield  {author} {\bibinfo {author} {\bibfnamefont {S.}~\bibnamefont
  {Keren-Zur}}, \bibinfo {author} {\bibfnamefont {O.}~\bibnamefont {Avayu}},
  \bibinfo {author} {\bibfnamefont {L.}~\bibnamefont {Michaeli}}, \ and\
  \bibinfo {author} {\bibfnamefont {T.}~\bibnamefont {Ellenbogen}},\
  }\href@noop {} {\bibfield  {journal} {\bibinfo  {journal} {ACS Photonics}\
  }\textbf {\bibinfo {volume} {3}},\ \bibinfo {pages} {117} (\bibinfo {year}
  {2015})}\BibitemShut {NoStop}%
\bibitem [{\citenamefont {Krasnok}\ \emph {et~al.}(2017)\citenamefont
  {Krasnok}, \citenamefont {Tymchenko},\ and\ \citenamefont
  {Al{\`u}}}]{krasnok2017nonlinear}%
  \BibitemOpen
  \bibfield  {author} {\bibinfo {author} {\bibfnamefont {A.}~\bibnamefont
  {Krasnok}}, \bibinfo {author} {\bibfnamefont {M.}~\bibnamefont {Tymchenko}},
  \ and\ \bibinfo {author} {\bibfnamefont {A.}~\bibnamefont {Al{\`u}}},\
  }\href@noop {} {\bibfield  {journal} {\bibinfo  {journal} {arXiv preprint
  arXiv:1706.07563}\ } (\bibinfo {year} {2017})}\BibitemShut {NoStop}%
\bibitem [{\citenamefont {Bond}(1965)}]{bond1965measurement}%
  \BibitemOpen
  \bibfield  {author} {\bibinfo {author} {\bibfnamefont {W.}~\bibnamefont
  {Bond}},\ }\href@noop {} {\bibfield  {journal} {\bibinfo  {journal} {Journal
  of Applied Physics}\ }\textbf {\bibinfo {volume} {36}},\ \bibinfo {pages}
  {1674} (\bibinfo {year} {1965})}\BibitemShut {NoStop}%
\bibitem [{\citenamefont {Shoji}\ \emph {et~al.}(1997)\citenamefont {Shoji},
  \citenamefont {Kondo}, \citenamefont {Kitamoto}, \citenamefont {Shirane},\
  and\ \citenamefont {Ito}}]{shoji1997absolute}%
  \BibitemOpen
  \bibfield  {author} {\bibinfo {author} {\bibfnamefont {I.}~\bibnamefont
  {Shoji}}, \bibinfo {author} {\bibfnamefont {T.}~\bibnamefont {Kondo}},
  \bibinfo {author} {\bibfnamefont {A.}~\bibnamefont {Kitamoto}}, \bibinfo
  {author} {\bibfnamefont {M.}~\bibnamefont {Shirane}}, \ and\ \bibinfo
  {author} {\bibfnamefont {R.}~\bibnamefont {Ito}},\ }\href@noop {} {\bibfield
  {journal} {\bibinfo  {journal} {JOSA B}\ }\textbf {\bibinfo {volume} {14}},\
  \bibinfo {pages} {2268} (\bibinfo {year} {1997})}\BibitemShut {NoStop}%
\end{thebibliography}
%

\end{document}